%% file: main.tex
\documentclass[manuscript,screen]{acmart}

\usepackage{graphicx}
\usepackage{subcaption}
\usepackage{wrapfig}
\usepackage{tikz}
\usepackage{multirow}
\begin{document}

\title{FTPrimitiveBench: A Benchmark Suite For Logical Computation Under Hardware-Motivated and Biased Noise Models}

\author{Shuwen Kan}
\email{sk107@fordham.edu}
\orcid{0009-0004-0726-5260}
\affiliation{%
  \institution{Physical and Computational Sciences, Pacific Northwest National Laboratory}
  \city{Richland}
  \state{Washington}
  \country{USA}
}
\affiliation{%
  \institution{Department of Computer and Information Sciences, Fordham University}
  \city{New York}
  \state{New York}
  \country{USA}
}

\author{Adrian Harkness}
\email{adrian.harkness@pnnl.gov}
\orcid{0009-0001-5518-6442}
\affiliation{%
  \institution{Physical and Computational Sciences, Pacific Northwest National Laboratory}
  \city{Richland}
  \state{Washington}
  \country{USA}
}
\affiliation{%
  \institution{Industrial and Systems Engineering, Lehigh University}
  \city{Bethlehem}
  \state{Pennsylvania}
  \country{USA}
}

\author{Zefan Du}
\email{zdu19@fordham.edu}
\orcid{0009-0005-6547-9168}
\affiliation{%
  \institution{Department of Computer and Information Sciences, Fordham University}
  \city{New York}
  \state{New York}
  \country{USA}
}

\author{Rod Rofougaran}
\email{rod.rofougaran@pnnl.gov}
\orcid{0000-0002-4747-7223}
\affiliation{
  \institution{Physical and Computational Sciences, Pacific Northwest National Laboratory}
  \city{Richland}
  \state{Washington}
  \country{USA}
}

\author{Sean Garner}
\email{sean.garner@pnnl.gov}
\orcid{}
\affiliation{
  \institution{Physical and Computational Sciences, Pacific Northwest National Laboratory}
  \city{Richland}
  \state{Washington}
  \country{USA}
}

\author{Chenxu Liu}
\email{chenxu.liu@pnnl.gov}
\orcid{0000-0003-2616-3126}
\affiliation{%
  \institution{Physical and Computational Sciences, Pacific Northwest National Laboratory}
  \city{Richland}
  \state{Washington}
  \country{USA}
}

\author{Ying Mao}
\email{ymao41@fordham.edu}
\orcid{0000-0002-4484-4892}
\affiliation{%
  \institution{Department of Computer and Information Sciences, Fordham University}
  \city{New York}
  \state{New York}
  \country{USA}
}

\author{Samuel Stein}
\email{samuel.stein@pnnl.gov}
\orcid{0000-0002-2655-8251}
\affiliation{%
  \institution{Physical and Computational Sciences, Pacific Northwest National Laboratory}
  \city{Richland}
  \state{Washington}
  \country{USA}
}


\newcommand{\sol}{FTPrimitiveBench}

\begin{abstract}

The pursuit of fault-tolerant quantum computing has highlighted the necessity of understanding how various error-correcting codes and their associated logical operations perform across diverse underlying physical systems. Such evaluation is typically carried out via noisy stabilizer simulation of a logical computation encoded by the scale of thousands of qubits and gates, combined with a noise model to evaluate its logical error rate on HPC-scale workload for relevant code distances and circuit depths. While the uniform depolarizing noise model remains a widely adopted standard, serving as a useful baseline approximation, its homogeneous assumptions are often insufficient to capture the complex and biased noise characteristics of physical hardware. In practice, heterogeneity, asymmetries, and correlations complicate the commonly assumed weak coupling between Pauli channel errors, measurement processes, and spatio-temporal error distributions. At the same time, these asymmetric and structured noise features can offer substantial opportunities when quantum error correction protocols and hardware are jointly designed and optimized. This motivates the need for models that more closely align with target hardware, while maintaining tractable simulation, thereby enabling more accurate performance estimation and more effective architectural design.

In this work, we introduce \sol, a systematic benchmarking approach for studying how key logical primitives interact with varying hardware-motivated noise assumptions. \sol~supports both custom specifications and representative structured noise families, including Pauli-noise bias, measurement-biased noise, and spatial or spatio-temporal non-uniformity, together with generators for a core set of surface-code Clifford primitives: logical memory, lattice surgery, logical Hadamard via transversal gate, and the logical phase (S) gate via lattice surgery.

Using \sol{}, we show that structured noise affects logical primitives in qualitatively distinct ways. Asymmetric patches substantially improve logical memory and lattice surgery under $Z$-biased noise, but this advantage diminishes for the transversal Hadamard, which exchanges $X$ and $Z$ channels mid-circuit and is therefore not bias-preserving. Under measurement-biased noise, lattice-surgery performance is sensitive to the number of syndrome-extraction rounds, highlighting the importance of temporal redundancy. Under non-uniform noise, the relative logical error rate(LER) stays close to the uniform-depolarizing baseline across the variance levels and code distances studied, given decoder priors matched to the underlying per-component error rates, indicating that the surface code is largely robust to spatial and spatio-temporal heterogeneity. We further observe that correlated minimum-weight perfect matching offers a clear advantage over its uncorrelated counterpart under depolarizing noise, while this gap narrows as the channel becomes strongly $Z$-biased.

These results complement homogeneous logical-memory benchmarks by extending analysis to active logical computation, where the interaction between noise structure and primitive implementation becomes important. By standardizing the relationship between noise-model specification and primitive construction, \sol{} enables reproducible comparative studies of QEC protocols and decoder performance, progressing towards connecting hardware characterization with logical-level performance analysis and providing a practical basis for hardware-aware co-design of fault-tolerant quantum architectures. \sol{} is fully opensourced and maintained on \href{https://github.com/ShuwenKan/FTPrimitiveBench}{GitHub}.
\end{abstract}





\pagestyle{plain}     
\maketitle
\thispagestyle{plain} 
\markboth{}{}         

\input{sections/introduction}
\input{sections/background}
\input{sections/related_work}

\input{sections/method}

\input{sections/Primitives}
\input{sections/evaluation}
\input{sections/discussion}

\section*{Acknowledgments}
This material is based upon work supported by the U.S. Department of Energy, Office of Science, National Quantum Information Science Research Centers, Quantum Science Center (QSC). This research was supported by PNNL’s Quantum Algorithms and Architecture for Domain Science (QuAADS) Laboratory Directed Research and Development (LDRD) Initiative.  The Pacific Northwest National Laboratory is operated by Battelle for the U.S. Department of Energy under Contract No. DE-AC05-76RL01830. This research used resources of the Oak Ridge Leadership Computing Facility, which is a DOE Office of Science User Facility supported under Contract No. DE-AC05-00OR22725. This research used resources of the National Energy Research Scientific Computing Center (NERSC), a U.S. Department of Energy Office of Science User Facility located at Lawrence Berkeley National Laboratory, operated under Contract No. DE-AC02-05CH11231. This research was supported in part by the National
Science Foundation (NSF) under grant agreements 2329020 and 2343535.

\bibliographystyle{ACM-Reference-Format}
\bibliography{ref}

\appendix

\end{document}

%% file: sections/introduction.tex
\section{Introduction}

Quantum error correction (QEC) is proving essential for the realization of large-scale, reliable quantum computers~\cite{beverland2022assessing, babbush2021focus, gidney2025factor}. By encoding logical qubits across many physical qubits, QEC protocols enable exponential error suppression~\cite{steane1996error, shor1996fault}. Recent experimental milestones in superconducting circuits~\cite{google2023suppressing, google_below_threshold}, trapped ions~\cite{berthusen2408experiments}, and neutral-atom arrays~\cite{bluvstein2024logical} have demonstrated the realization of logical operations and further bolstered motivation towards realising quantum error correction as the building block towards fault-tolerant quantum computing, whilst underscoring the importance of understanding its performance under realistic, hardware-dependent noise conditions.

In practice, noise in quantum hardware is inherently heterogeneous and often exhibits asymmetry, reflecting variations in device calibration, operation times, connectivity constraints, and dominant error mechanisms across platforms. Therefore, logical primitive behavior is closely coupled to the underlying physical noise, as shown in recent surface-code experiments on superconducting hardware~\cite{vezvaee2025surface}. This motivates systematic benchmarking approaches that enable controlled exploration of how heterogeneous and asymmetric noise structures influence logical performance. Such studies are typically carried out through large-scale simulations built on efficient stabilizer-circuit simulators~\cite{aaronson2004improved, gidney2021stim}, run on HPC infrastructure to reach statistically meaningful logical error rates, especially for circuits targeting high code distance and low physical error rate that comprise thousands of qubits and gates. Crucially, the insights obtained from these studies depend on how well the assumed noise models capture the relevant structure and variability of the target hardware.

Among commonly used noise models, uniform depolarizing noise—where every fault location is assigned the same error rate $p$—has served as a standard baseline for comparison~\cite{fowler2012proof, fowler2012surface, wang2011surface, wang2009threshold, tomita2014low}. A step towards more hardware-informed noise model is the SI1000 noise model~\cite{gidney2021fault}, which incorporates operation-dependent error rates motivated by superconducting platforms. While these models have been instrumental in advancing the study of quantum error correction, emerging experimental capabilities~\cite{google_below_threshold, cong2022hardware, putterman2025hardware} suggest incorporating richer and more hardware-aligned noise models may offer additional opportunities for understanding and improving logical performance. In particular, capturing features such as Pauli bias~\cite{aliferis2009fault, lescanne2020exponential, putterman2025hardware},  measurement-dominated regimes in which readout infidelity substantially exceeds gate error rates~\cite{google_below_threshold, bluvstein2024logical, krinner2022realizing}, and spatial or temporal non-uniformity~\cite{moses2023race, wilen2021correlated, mcewen2022resolving, klimov2018fluctuations} can provide a more faithful reflection of hardware behavior while enabling more informative comparisons across protocols. Accounting for such heterogeneity can materially shape logical performance and, in some cases, refine the conclusions drawn from uniform noise models~\cite{PhysRevResearch.6.043249}, making the interaction between structured noise and protocol-level variation an increasingly important research direction.

In this work, we present \sol{}, a systematic benchmarking methodology for evaluating QEC primitives under hardware-motivated noise structures. \sol{} establishes a flexible interface between noise-model specification and circuit generation, exposing parameter assignment at four levels of granularity---global, per-component, per-round, and per-(component, round). A fully customized noise model thereby captures both a hardware calibration profile and the physical-overhead errors induced by compilation: gate and SPAM rates are sourced from the device profile, while idle channels are accumulated from the compiled syndrome-extraction circuit via gate duration and $T_1$/$T_2$ parameters, factoring circuit-level idle exposure into the noise budget rather than relying on nominal gate-error rates alone. The contributions are:

\begin{itemize}
    \item \textbf{Flexible Noise Modeling:} Three usage modes built on this interface: pre-packaged structured-noise families (Pauli bias, measurement bias, spatial/spatio-temporal non-uniformity) for controlled comparative studies, selective per-qubit and per-round overrides for device heterogeneity and drift, and fully customized hardware-aligned specifications.
    \item \textbf{Surface-Code Primitive Generation:} Automated, high-level circuit generators for the core logical Clifford primitives: logical memory, lattice surgery, transversal Hadamard, and the lattice-surgery-based phase ($S$) gate, exported as \textsc{Stim} circuits with detector annotations and logical observables, so outputs drop directly into existing simulation and decoding pipelines.
    \item \textbf{Standardized Benchmarking Pipeline:} A reproducible workflow that pairs noise-model specification with primitive generation through a single Stim-format output, enabling direct comparative studies of decoders, simulators and primitives under matched, hardware-motivated noise assumptions.
\end{itemize}

Applied to the rotated surface code, \sol{} reveals that the same hardware-motivated noise structure produces qualitatively distinct logical behavior across primitives. Asymmetric patches substantially reduce the logical error rate of memory and lattice surgery under $Z$-biased Pauli noise, but the transversal Hadamard exchanges $X$ and $Z$ channels mid-circuit and dissolves that geometry-tailored advantage. Under measurement-biased noise, the round count minimizing the lattice-surgery LER drifts upward with the bias factor, foregrounding temporal redundancy as an architectural knob invisible to uniform-depolarizing analyses. Non-uniform noise, by contrast, leaves the relative LER close to the uniform-depolarizing baseline across the variance levels and code distances surveyed once decoder priors match the realized per-component rates, indicating that the rotated surface code is largely robust to per-qubit calibration heterogeneity. The correlated-MWPM~\cite{fowler2013optimal} advantage over its uncorrelated counterpart~\cite{higgott2025sparse}is largest under depolarizing noise and narrows as the channel becomes strongly $Z$-biased, evidencing that decoder choice is itself shaped by the underlying error structure.

The built-in families recover the uniform-depolarizing baseline and SI1000-style operation-dependent assignment as special cases, while the per-component and per-round override paths expose the finer-grained structural axes that contemporary hardware exhibits. While \sol{} is architected for general extensibility to a wide range of codes, this work focuses on a comprehensive evaluation of surface-code primitives given their central role in current hardware roadmaps. \sol{} serves as a practical tool for the hardware--software co-design of fault-tolerant systems, providing the infrastructure to optimize logical-level performance for the diverse and evolving signatures of modern quantum platforms.

The remainder of this paper is organized as follows. Section~\ref{sec:background} provides background on quantum error correction. Section~\ref{sec:related} positions \sol{} against prior simulators, primitive generators, noise models, and benchmark suites. Section~\ref{sec:primitivebench} presents the architecture of \sol{}. Section~\ref{sec:evaluation} demonstrates the utility of the framework through an evaluation of surface-code primitives under structured noise families, and Section~\ref{sec:discussion} discusses extensibility and implications for hardware--software co-design.

%% file: sections/background.tex
\section{Background}
\label{sec:background}
This section provides the background for \sol, beginning with the principles of quantum error correction and the surface code, followed by the construction of logical primitives for computation, and the simulation methods used to study their performance.

\subsection{Quantum Error Correction}

Quantum error correction (QEC) protects quantum information by encoding logical information into a larger Hilbert space. An $[n, k, d]$ QEC code encodes $k$ logical qubits into $n$ physical data qubits, and can correct up to $\lfloor (d-1)/2\rfloor$ errors. QEC codes are most often described via stabilizer formalisms ~\cite{gottesman1997stabilizer}, where the encoded logical space is defined by the simultaneous $+1$ eigenspace over a set of stabilizers. Stabilizers are defined by Pauli strings $P$, where $P = \bigotimes_{j=1}^n P_j$ and $ P_j \in \{I,X,Y,Z\}$. A code $\mathcal{C}$ can be defined over a set of stabilisers $S_\mathcal{C} = \{\, |\psi\rangle \mid S_i |\psi\rangle = |\psi\rangle,\ \forall i \,\}.$
The codes stabilizers can be implemented in hardware via the following procedure.


\begin{enumerate}
    \item \textbf{Entangling Step:} A syndrome-extraction circuit applies a sequence of entangling gates - typically CNOT or CZ gates - between an ancilla qubit and the data qubits involved in a specific stabilizer check. These gates are executed according to a carefully designed schedule~\cite{kishony2026surface, fujiu2025dense, hirai2026no} to prevent hardware scheduling conflicts and to mitigate the propagation of highly consequential error channels such as "hook errors".
    
    \item \textbf{Measurement and Detection:} Following the entangling operators, the ancilla is measured to extract the eigenstate of the associated stabilizer. Since measurements are error prone, the absolute value of a single measurement is often less important than a sequence of detection events. A detector is defined as the sum of two sequential measurements over $\mathbb{F}_2$, that should be deterministically 0 (e.g., yielding a parity of 0) in the absence of noise. Conversely, a sum of +1 indicates that an unexpected change has occurred, signaling that one or more errors have affected the system.
    
    \item \textbf{Syndrome Collection:} This process is repeated for many rounds, producing a measurement record for each stabilizer. The resulting data forms a three-dimensional "space-time volume." This structure consists of two spatial dimensions derived from the physical qubit layout and one temporal dimension representing the repeated rounds of measurement. This is then passed to the decoder to decode the most likely history of errors, which are to be corrected. 
\end{enumerate}

A classical decoding algorithm parses the detector information, analyzes the detection events and produces a suitable correction consistent with the observed syndromes. The decoder’s effectiveness depends heavily on its prior knowledge of the hardware's noise profile. In statistical decoding, such as Minimum-Weight Perfect Matching (MWPM)~\cite{higgott2022pymatching}, the decoder assigns weights to potential error paths based on the probability of specific physical errors. A mismatch between these internal priors and the actual hardware noise can lead to sub-optimal matching and increased logical error rate. Since the decoder's performance is tightly coupled with the noise model it assumes, a growing body of research focuses on optimising these decoder priors through hardware-aware calibration and automated noise-model estimation~\cite{sivak2024optimization, wang2023dgr, sivak2025reinforcement}. 

Physical data qubits are often not corrected through real-time physical gates, as such gates are themselves noisy and may introduce additional errors. Instead, error correction is performed in software. The classical control system maintains a Pauli frame - a classical record of all detected flips - to track the cumulative Pauli frame of the logical qubit. These accumulated errors are then accounted for during the final measurement of the quantum algorithm, where the classical computer adjusts the outcomes to ensure the final output is logically correct.

\subsection{Surface Code}

\begin{figure}
    \centering
    \includegraphics[width=1\linewidth]{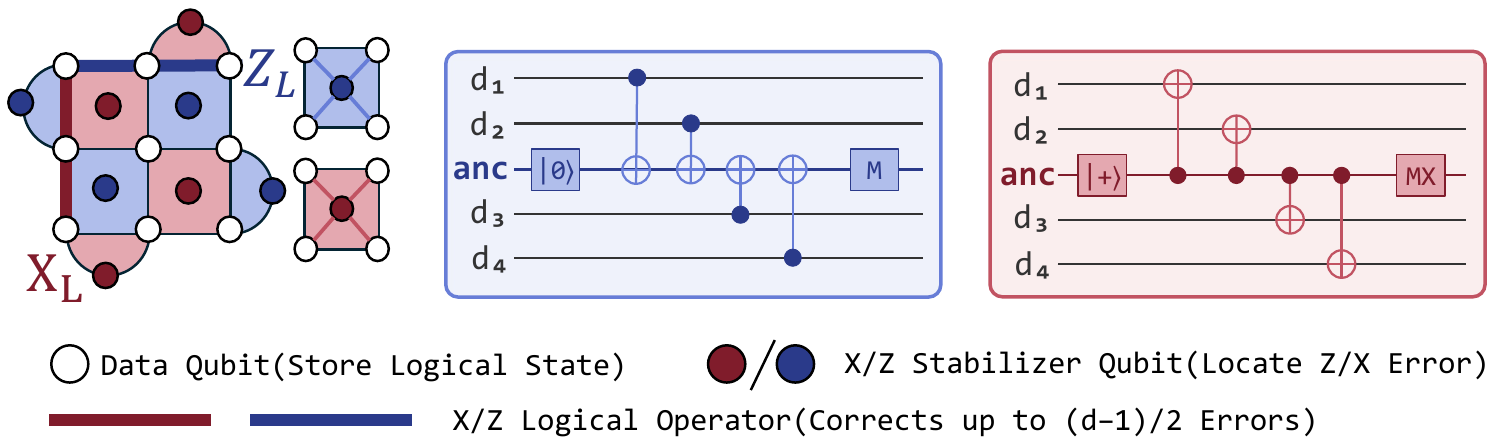}
    \caption{Illustration of a surface-code logical qubit. Left: a distance-3 patch showing data qubits (white) and interleaved $X$- and $Z$-stabilizer ancilla qubits (red/blue) used to detect errors; logical operators $X_L$ and $Z_L$ correspond to strings spanning the patch. Middle: syndrome extraction circuit for $Z$-stabilizers using an ancilla prepared in $|0\rangle$, entangled with neighboring data qubits and measured in the computational basis. Right: syndrome extraction circuit for $X$-stabilizers using an ancilla prepared in $|+\rangle$ and measured in the $X$ basis. Repeated stabilizer measurements enable detection and correction of errors up to $(d-1)/2$.}
    \label{fig:surface-code-overview}
\end{figure}

The surface code~\cite{kitaev2003fault, fowler2012surface} is among the most widely studied QEC codes, with implementations demonstrated across multiple hardware platforms~\cite{google_below_threshold, google2023suppressing, zhao2022realization, krinner2022realizing, bluvstein2024logical, erhard2021entangling, berthusen2408experiments, bluvstein2025architectural}. We focus on the rotated variant~\cite{horsman2012surface} for its favorable encoding rate. The surface code is well suited to near-term hardware due to its high error-rate threshold, local connectivity requirements, and efficient decoding. Figure~\ref{fig:surface-code-overview} illustrates the rotated surface code patch layout and the syndrome-extraction circuits used to measure the $X$- and $Z$-type stabilizers discussed in the remainder of this section.

\subsubsection{Code Construction}
In the rotated surface code~\cite{horsman2012surface, fowler2009high, fowler2012surface}, physical qubits are arranged on a 2D square lattice. The code is defined by stabilizer generators: weight-4 operators ($XXXX$ or $ZZZZ$) in the interior and weight-2 operators ($XX$ or $ZZ$) along the boundaries, arranged in a checkerboard pattern to detect bit-flip ($X$) and phase-flip ($Z$) errors. The logical operators $X_L$ and $Z_L$ correspond to chains of Pauli operators spanning the lattice between opposite boundaries.

From a decoding perspective, error recovery maps naturally onto the matching problem in graph theory. MWPM~\cite{fowler2013optimal, higgott2022pymatching, wu2023fusion, Higgott2025sparseblossom} treats detection events as nodes and potential faults as weighted edges, identifying the minimum-weight matching to reconstruct the most probable error chains.

\subsubsection{Logical Primitives}

\sol~focuses on logical Clifford-group operations that serve as building blocks for fault-tolerant surface-code architectures. For detailed circuit constructions, we refer readers to~\cite{fowler2018low, bombin2023logical, Gidney2024inplaceaccessto}.

\begin{itemize}
    \item \textbf{Logical Memory}: Identity operation, preserving a logical state over $t$ syndrome-extraction rounds. This primitive is used to characterize the ability of the code to stabilize a logical state over time, including the effects of data-qubit noise, gate errors, and imperfect measurements.
    
    \item \textbf{Lattice Surgery}: Realizes entangling operations between surface-code patches through joint logical parity measurements such as $X_LX_L$ or $Z_LZ_L$~\cite{horsman2012surface, fowler2018low}. Patches are temporarily merged along a shared boundary, requiring only nearest-neighbor physical interactions.


    \item \textbf{Logical Hadamard ($H_L$)}: Implemented transversally by applying a physical Hadamard to every data qubit, which results in the exchange of $X$-type and $Z$-type stabilizers. Additional code-deformation or routing steps may be needed to restore the patch orientation~\cite{huggins2025fluid, bombin2023logical}.


    \item \textbf{Logical S gate ($S_L$)}: Not transversal in the rotated surface code; typically realized via teleportation using a logical $|Y\rangle$ resource state~\cite{Gidney2024inplaceaccessto}, comprising lattice surgery.
\end{itemize}


Logical Pauli gates ($X_L, Y_L, Z_L$) require no physical gate execution and are tracked classically through Pauli-frame updates.

\subsubsection{Universal Fault-Tolerant Computation with Magic States}

Clifford gates alone are insufficient for universal quantum computation. To perform universal quantum computation, at least one non-Clifford gate must be supported ~\cite{harkness2026ftcircuitbench}. A common computational model is is Clifford+T~\cite{ross2014optimal} or Pauli-based computation~\cite{litinski2019game}. In the surface code, non-Clifford gates can be realised via gate teleportation over a magic state - such as $|T\rangle$ or $|CCZ\rangle$ - through lattice surgery.  Preparation of high-fidelity magic states can be realized via distillation or cultivation \cite{litinski2019magic,gidney2024magic}, and recent cultivation protocols~\cite{gidney2024magic} have significantly reduced the associated overhead, highlighting the need for joint optimisation of both Clifford and non-Clifford computation.

\subsubsection{Compilation and Spacetime Overhead}
 
Compiling a surface-code program involves mapping quantum circuits into a three-dimensional arrangement of logical primitives, often via intermediate representations such as the ZX-calculus~\cite{van2020zx, de2020zx} or Pauli-based computation~\cite{litinski2019game}. The resulting logical computation, expressed as hardware-level instructions like those in~\cite{gidney2025factor, tan2024sat}, can be viewed as a spacetime diagram—two spatial dimensions for the physical qubit footprint and one temporal dimension for syndrome-extraction rounds—whose volume is a common optimization target. Since these structures are topologically flexible, a logical algorithm can be reshaped by merging, splitting, or rerouting logical operations, provided the underlying topology is preserved~\cite{zhou2026topols, tan2024sat}. Resource-estimation frameworks such as FLASQ~\cite{huggins2025fluid} quantify costs in units of $d^3$.


\subsection{Quantum Error-Correction Simulation}

\subsubsection{Stabilizer Simulation}

Simulation is a central tool for evaluating fault-tolerant primitives. A logical primitive is compiled into a noisy syndrome-extraction circuit, physical faults are sampled from a noise model, and the resulting detection events are decoded to estimate the probability of logical failure. This Monte Carlo approach is the standard methodology for benchmarking QEC protocols~\cite{gidney2021stim, garner2025stabsim}.

Clifford circuits can be simulated efficiently within the stabilizer formalism using a compact tableau representation~\cite{aaronson2004improved}, avoiding the exponential cost of state-vector simulation. High-performance simulators such as Stim~\cite{gidney2021stim} enable large-scale Monte Carlo trials, and the widespread adoption of Stim's circuit syntax and Detector Error Model (DEM) format has established a practical standard for interoperability across the QEC simulation ecosystem~\cite{garner2025stabsim,garner2025stabsim}.

\subsubsection{Noise Models}

The relevance of a simulated logical error rate depends on the underlying noise model. Three levels of abstraction are common in the QEC literature~\cite{kitaev2003fault, dennis2002topological,wang2003confinement, wang2011surface, wang2009threshold, stephens2014fault,fowler2012surface, stephens2014fault, gidney2021stim}:
    
\begin{itemize}

    \item \textbf{Code Capacity Noise:} stochastic Pauli errors are applied to the data qubits, with no measurement error, useful for evaluating code properties, and modeling of the codes error propagation statistics. \cite{kitaev2003fault, dennis2002topological}.
    
    \item \textbf{Phenomenological Noise:} Extends the code-capacity model by adding independent measurement errors. This provides the baseline for studying code properties and thresholds ~\cite{wang2003confinement, wang2011surface, wang2009threshold, stephens2014fault}.

    \item \textbf{Circuit-Level Noise:} Injects error channels after every physical operation, including single- and two-qubit gates, resets, measurements, and idle periods. This model can be tightly coupled to approximate the noise structure of an architecture. This allows for modelling of error propagation such as "hook errors" and time-correlated faults, providing a more hardware-realistic basis for benchmarking~\cite{fowler2012surface, stephens2014fault, gidney2021stim}.

\end{itemize}

\subsubsection{Decoding}

Decoding is the classical process of parsing and processing syndrome data to infer and correct physical errors. The most widely used decoder for the surface code is MWPM such as sparse blossom implemented in PyMatching~\cite{higgott2022pymatching, higgott2025sparse} and micro blossom\cite{wu2025micro}. Other approaches include maximum-likelihood decoders such as Tesseract~\cite{beni2025tesseractdecoder}, tensor network decoder~\cite{piveteau2024tensor} and belief-propagation deocder variants~\cite{roffe_decoding_2020, Roffe_LDPC_Python_tools_2022}. Standardized benchmarking efforts such as decoder-bench~\cite{maurya2025decoder} provide platforms for comparing decoder accuracy, speed, and scalability.

%% file: sections/related_work.tex
\section{Related Work}
\label{sec:related}
\sol{} draws on an active ecosystem of QEC simulators, decoders, and noise-model proposals, and aims to complement them by linking hardware-motivated noise specification with logical-primitive evaluation. We position the contribution against four threads of prior work below.

\textbf{Stabilizer simulators.} \textsc{Stim}~\cite{gidney2021stim} and other tools (\textsc{StabSim}~\cite{garner2025stabsim}, \textsc{tsim}~\cite{tsim2026}) provide highly optimized stabilizer circuit simulation and sampling, with Stim additionally defining the Detector Error Model (DEM) format that has become a common interchange across the ecosystem. These tools operate on circuits that are already constructed: they consume a fully-specified Stim program with detectors and observables wired in, and are agnostic to how that program was generated. \sol{} sits upstream of this layer, emitting standards-conforming Stim circuits with detectors and observables wired through every phase of the chosen primitive, so the existing simulation and decoding pipeline can be reused without modification.

\textbf{Primitive and computation generators.} A complementary line of work focuses on constructing the circuits themselves. The Stim toolchain~\cite{gidney2021stim, gidney_crumble} ships example circuits for common QEC primitives, and \texttt{tqec}~\cite{suau2026tqec} is a design-automation framework that compiles logical computations on the surface code from 3D pipe diagrams into Stim circuits, targeting end-to-end compilation of arbitrary topological computations from a high-level specification. Its circuit-level outputs can be re-instantiated under the structured-noise interface developed here, allowing the two layers to be combined. In addition, a number of individual primitives have been published as part of standalone studies~\cite{gidney2024magic, Gidney2024inplaceaccessto, shalby2025optimized, chen2026efficient, chen2026transversal}, each adopting its own circuit-construction conventions, noise placement, and parameterization. \sol{} takes a different design point, pairing a fixed primitive set with a single noise-model interface so that one structured-noise specification can be reused across primitives. The shared output format also lets this interface drive circuits emitted by either the Stim toolchain or \texttt{tqec}.

\textbf{Hardware-motivated noise models.} Two complementary threads of prior work shape this space. On the software side, \texttt{tqec}~\cite{suau2026tqec} ships built-in noise models that primarily target global, uniform parameter assignment, providing well-tested baselines such as uniform depolarizing and SI1000-style channels; the conventions used in standalone primitive studies such as~\cite{Gidney2024inplaceaccessto} follow the same global pattern. \sol{} aims to broaden this surface to per-component, spatial, and spatio-temporal granularity within the same interface. On the modeling side, SI1000~\cite{gidney2021fault} encodes operation-time asymmetries representative of superconducting platforms, while biased-noise studies on neutral-atom platforms~\cite{cong2022hardware} motivate $Z$-biased Pauli channels that bias-tailored codes such as XZZX~\cite{tuckett2018ultrahigh, tuckett2019tailoring} are designed to exploit. Non-uniformity~\cite{auger2017fault, palmer2025boundaries} and probabilistic per-qubit error-rate variation~\cite{tiurev2023correcting, sriram2024non} perturb thresholds and, under strong heterogeneity, can alter the logical-error scaling. \sol{} keeps these two choices separable: each of the structured-noise families above corresponds to a configuration of the same noise-model interface (\S\ref{sec:primitivebench}) and applies in the same way to every supported primitive, so a noise feature studied here in the context of memory can also be examined against lattice surgery or the phase gate without rewriting the circuit generator.

\textbf{Other fault-tolerance benchmarks.} \textsc{decoder-bench}~\cite{maurya2025decoder} targets the decoding stage, supplying code-capacity, phenomenological, and circuit-level noise models for decoder evaluation. FTCircuitBench~\cite{harkness2026ftcircuitbench} sits above the primitive layer, handling algorithm transpilation onto the primitives benchmarked here and reporting algorithm-level statistics such as $T$ count and Clifford count. \sol{} is intended to complement these efforts at the primitive layer, where cross-primitive comparison under matched hardware assumptions can be expressed as a configuration change in noise models.

%% file: sections/method.tex
\section{\sol}
\label{sec:primitivebench}

In this work, we introduce \sol, a systematic approach for benchmarking logical primitives in the surface code under hardware-realistic noise. \sol unifies noise-model specification with stabilizer-circuit representations of logical computation, enabling controlled and reproducible evaluation of logical error rates across diverse noise profiles, hardware assumptions, and decoding strategies. It provides (1) a flexible noise-model interface that decouples physical noise specification from circuit generation, accompanied by four built-in hardware-motivated families that extend beyond uniform depolarizing noise; and (2) a high-level generator for surface-code logical primitives spanning the logical Clifford group, including memory, lattice surgery, transversal Hadamard, and the lattice-surgery-based phase gate.

By enabling consistent comparisons across noise regimes and primitives, \sol makes it possible to study how logical performance depends jointly on noise structure, circuit implementation, and decoding. This reveals how bias, asymmetry, and non-uniformity shape logical error rates across different primitives. By making these interactions explicit, \sol provides a principled basis for comparing fault-tolerant architectures and for guiding hardware–software co-design.

\paragraph{Motivation}
With the maturation of fault-tolerant quantum computation (FTQC), the focus is shifting from establishing feasibility to understanding and optimizing performance under realistic conditions. In this setting, it becomes increasingly important to characterize how performing actual logical computation and the physical noise model interact, and their respective performance. Since conclusions about logical performance can depend materially on details of underlying assumptions, it is important to be able to benchmark and explore the infinitely large design space of driving logical computation.

In practice, physical noise is not uniform. Error rates vary between qubits, gate types, measurements, spatial locations, correlated errors, and time~\cite{klimov2018fluctuations,mcewen2022resolving, evered2023high,moses2023race, wilen2021correlated}. As the performance of fault tolerant architectures is closely tied to the underlying noise model, how computation is compiled and realised, and the decoder, the exploration and interplay of these properties is well motivated. 

Simultaneously, fair comparison across QEC proposals remains difficult as simulation studies do not always adopt standardized modeling assumptions. Even within the commonly used uniform depolarizing setting, prior work differs in important implementation details; for example, some studies include idling errors~\cite{gidney2021fault}, while others do not~\cite{marton2025lattice}. As a result, reported logical error rates and threshold-related trends are often not directly comparable across the literature. An important step beyond uniform depolarizing noise is SI1000~\cite{gidney2021fault}, a superconducting-inspired noise model that assigns operation-dependent error rates using a single base parameter $p$; for example, single-qubit gates are scaled as $p/10$, two-qubit gates as $p$, and measurement operations as $5p$. Although SI1000 provides a more structured and hardware-motivated alternative to uniform depolarizing noise, it still captures only a limited portion of the broader space of heterogeneous noise processes relevant to QEC research. 

These limitations motivate the need for a standardized and reproducible benchmarking framework that enables systematic study of noise heterogeneity across logical primitives, codes, and architectures. Such a framework is valuable not only for fair comparison, but also for supporting controlled investigation of how different forms of structured noise affect logical performance.

\subsection{Noise Model Interface}
\label{sec:noise-interface}

A central design goal of \sol{} is to decouple logical-circuit generation from physical noise specification. Starting from a noiseless stabilizer circuit, \sol~injects stochastic error channels according to a user-specified noise description, enabling the same logical primitive to be evaluated under a wide range of hardware assumptions. Noise is restricted to stochastic Pauli channels to remain compatible with efficient stabilizer simulation. 

Section~\ref{sec:noise-channels} specifies the per-operation noise channels (gate, SPAM, idle) and their parameterizations. Section~\ref{sec:noise-assignment} then describes how channel parameters are assigned to specific physical qubits, interactions, and syndrome-extraction rounds, supporting global, spatial, temporal, and spatio-temporal heterogeneity.

\subsubsection{Noise Channels}
\label{sec:noise-channels}

\sol{} distinguishes three classes of physical noise: gate errors, SPAM errors, and idling errors. For each class, the framework provides a default parameterization suitable for baseline studies, while also exposing the underlying channel parameters so that users can override the defaults with hardware-calibrated specifications.
\begin{itemize}
    \item \textbf{Gate errors.}
    Each physical gate is followed by a Pauli error channel. By default, gate errors are modeled as depolarizing noise, but this default can be replaced by a more general Pauli channel to capture bias or correlated faults.
    \begin{itemize}
        \item \textit{Single-qubit gates.}
        For a single-qubit gate on qubit \(i\), the most general supported channel is
        \begin{equation}
            \mathcal{P}^{(1)}_i(\rho)
            =
            (1-p^{(1)}_\Sigma)\rho
            + p^{(1)}_x X_i \rho X_i
            + p^{(1)}_y Y_i \rho Y_i
            + p^{(1)}_z Z_i \rho Z_i,
            \qquad
            p^{(1)}_\Sigma = p^{(1)}_x + p^{(1)}_y + p^{(1)}_z.
        \end{equation}
        The default single-qubit model is the depolarizing special case
        \(p^{(1)}_x = p^{(1)}_y = p^{(1)}_z = p^{(1)}/3\).
        Users may override this default by specifying the three Pauli components
        \(p^{(1)}_x,p^{(1)}_y,p^{(1)}_z\) independently, enabling, for example, biased single-qubit noise.

        \item \textit{Two-qubit gates.}
        For a two-qubit gate on qubits \(i,j\), the supported channel is
        \begin{equation}
            \mathcal{P}^{(2)}_{ij}(\rho)
            =
            (1-p^{(2)}_\Sigma)\rho
            +
            \sum_{P\in\{I,X,Y,Z\}^{\otimes 2}\setminus\{II\}}
            p^{(2)}_P\, P_{ij}\rho P_{ij},
            \qquad
            p^{(2)}_\Sigma = \sum_{P\neq II} p^{(2)}_P.
        \end{equation}
        The \(15\) parameters \(\{p^{(2)}_P\}\) capture both axis-dependent marginals
        (e.g., \(XI\), \(IZ\)) and correlated two-qubit faults (e.g., \(ZZ\), \(XY\)).
        The default two-qubit model is the depolarizing special case in which all
        \(p^{(2)}_P\) for \(P\neq II\) are equal.
        Users may override this default by specifying the full set of \(15\) Pauli probabilities.
    \end{itemize}

    \item \textbf{SPAM errors.}
    State-preparation and measurement errors are modeled separately from gate errors using basis-dependent scalar rates. For basis \(b\in\{X,Z\}\), the default SPAM channel is
    \begin{equation}
    \mathcal{P}^{(\mathrm{SPAM})}_{i,b}(\rho)
    =
    (1-p_b)\rho + p_b\,\sigma_b \rho \sigma_b,
    \end{equation}
    where \(p_b\) is the reset or measurement error rate in basis \(b\). Reset noise is applied immediately after preparation, and measurement noise immediately before readout. These default rates can be overridden with basis-specific calibrated values.

    \item \textbf{Idling errors.}
    \sol{} accounts explicitly for errors accumulated while qubits wait between operations.
    The circuit is resolved into discrete time ticks containing non-conflicting operations.
    For each tick, the simulator determines the tick duration as the longest gate duration among the active operations. A qubit that is not active for the full tick is assigned an idle interval equal to the remaining time in that tick.

    By default, idling noise is generated from the qubit-specific coherence parameters \(T_1\) and \(T_2\) using a Pauli-twirled approximation(PTA)~\cite{geller2013efficient, katabarwa2015logical} to amplitude damping and dephasing:
    \begin{align}
        p_x(t) &= p_y(t) = \tfrac{1}{4}\!\left(1-e^{-t/T_1}\right), \\
        p_z(t) &= \tfrac{1}{2}\!\left(1-e^{-t/T_2}\right)
                 - \tfrac{1}{4}\!\left(1-e^{-t/T_1}\right),
    \end{align}
    where \(t\) is the idle duration and \(T_2^{-1} = (2T_1)^{-1} + T_\phi^{-1}\).
    This yields the effective idling channel
    \begin{equation}
        \mathcal{P}^{\mathrm{idle}}_i(\rho)
        =
        p_0 \rho
        + p_x X_i \rho X_i
        + p_y Y_i \rho Y_i
        + p_z Z_i \rho Z_i,
        \qquad
        p_0 = 1 - p_x - p_y - p_z.
    \end{equation}
    Users may override this default by supplying calibrated idle-channel probabilities
    \(p_x(t), p_y(t), p_z(t)\), either globally or at the level of individual devices and idle windows.
\end{itemize}

As a result, the logical error rate depends not only on nominal gate-error parameters, but also on the detailed circuit schedule, operation durations, and time spent idling between gates~\cite{tomita2014low, beale2025logical}.

\subsubsection{Component and Round Assignment}
\label{sec:noise-assignment}

\sol{} attaches the channel parameters of Section~\ref{sec:noise-channels} to two kinds of hardware objects. A \emph{physical qubit} carries the error rates and durations of its single-qubit operations (gate, reset, readout) and its coherence parameters $T_1$ and $T_2$. A \emph{physical interaction}, indexed by a qubit pair, carries the error rate and duration of the two-qubit gate on that pair. Each such parameter is then bound at one of the following four levels of granularity:
\begin{itemize}
    \item \textbf{Global assignment:} a single parameter set applies to every qubit, every interaction, and every round, recovering homogeneous noise models such as the uniform-depolarizing baseline.
    \item \textbf{Spatial assignment:} parameters vary across qubits and interactions but remain fixed in time. This captures static device heterogeneity that persists for the duration of an experiment.

    \item \textbf{Temporal assignment:} parameters vary from round to round but are shared across components within a round, allowing the model to represent drift, fluctuations, or transient degradation over repeated syndrome-extraction cycles~\cite{klimov2018fluctuations,spitz2018adaptive}. Here, a round is defined as one stabilizer-extraction cycle, from ancilla reset through entangling operations to ancilla measurement.
    \item \textbf{Spatio-temporal assignment:} spatial and temporal variation can be combined, so that each component follows its own time-dependent noise profile.
\end{itemize}

This four-tier hierarchy lets a single circuit be re-evaluated under progressively richer noise environments without modifying the upstream generator: a calibration profile measured on hardware can be plugged in at the spatial or spatio-temporal level, while controlled studies can fall back to the global level for clean comparisons against published baselines.

\subsection{Built-in Noise Families}
\label{subsec:builtin_noise}

On top of the interface defined above, \sol{} provides four built-in noise families, summarized in Table~\ref{tab:builtin_families} and detailed in the subsections that follow. These serve as reusable defaults, but any subset of their parameters can still be overridden through the component- and round-level interface of Section~\ref{sec:noise-assignment}.

\begin{table*}[t]
\centering
\caption{Built-in noise families in \sol{}. Each column specifies the channel form attached to every operation class (rows) and the assignment pattern across components and rounds. The biased $X$/$Z$ family redistributes a fixed error budget $p$ along a chosen Pauli axis $A\in\{X,Z\}$ using a bias factor $\eta$. The measurement-biased family applies the uniform baseline to gates and idling and rescales only the SPAM rate by $\eta$. The non-uniform family is a meta-modifier on the uniform baseline: every rate $p$ is replaced by a Gaussian-perturbed rate $p_c = p(1+\delta_c)$ (clipped to $[0,0.5]$), where $c$ indexes physical components for the spatial variant and (component, round) pairs for the spatio-temporal variant. Detailed equations are given in the indicated subsections.}
\label{tab:builtin_families}
\renewcommand{\arraystretch}{1.45}
\setlength{\tabcolsep}{4pt}
\small
\begin{tabular}{@{}p{2cm} p{2.2cm} p{3.6cm} p{2.2cm} p{3.4cm}@{}}
\toprule
 & \textbf{Uniform} & \textbf{Biased $X$/$Z$} & \textbf{Meas.\ Bias} & \textbf{Non-Uniform} \\
 & (\S\ref{subsec:unifom}) & (\S\ref{subsec:biased_xz}) & (\S\ref{subsec:meas_biased_noise}) & (\S\ref{subsec:nonuniform_noise}) \\
\midrule
Parameterization
& $p$
& $p$, $\eta$, $A\in\{X,Z\}$
& $p$, $\eta$
& $p$, $\sigma$,\newline $\delta_c\sim\mathcal{N}(0,\sigma^2)$ \\
\midrule
1Q gate (Pauli)
& $p_X{=}p_Y{=}p_Z{=}\tfrac{p}{3}$
& $p_{A^\perp}{=}p_Y{=}\tfrac{p}{\eta+2}$,\newline $p_A{=}\tfrac{\eta p}{\eta+2}$
& $p_X{=}p_Y{=}p_Z{=}\tfrac{p}{3}$
& $p_X{=}p_Y{=}p_Z{=}\tfrac{p_c}{3}$,\newline $p_c{=}p(1{+}\delta_c)$ \\
\midrule
2Q gate (Pauli)
& $p_P{=}\tfrac{p}{15}$ for all $P{\neq}II$
& $p_P{=}\begin{cases}\tfrac{\eta p}{12+3\eta}, & P\in\{AI,IA,AA\}\\[2pt] \tfrac{p}{12+3\eta}, & \text{otherwise}\end{cases}$
& $p_P{=}\tfrac{p}{15}$ for all $P{\neq}II$
& $p_P{=}\tfrac{p_c}{15}$ for all $P{\neq}II$ \\
\midrule
SPAM (flip)
& $p_{\text{flip}}{=}p$
& $p_{M_A}{=}\tfrac{2p}{1+\eta}$,\newline $p_{M_{A^\perp}}{=}\tfrac{2\eta p}{1+\eta}$
& $p_{\text{flip}}{=}\eta p$
& $p_{\text{flip}}{=}p_c$ \\
\midrule
Idle (Pauli)
& $p_X{=}p_Y{=}p_Z{=}\tfrac{p}{3}$
& $p_{A^\perp}{=}p_Y{=}\tfrac{p}{\eta+2}$,\newline $p_A{=}\tfrac{\eta p}{\eta+2}$
& $p_X{=}p_Y{=}p_Z{=}\tfrac{p}{3}$
& $p_X{=}p_Y{=}p_Z{=}\tfrac{p_c}{3}$ \\
\midrule
Assignment
& global
& global
& global
& per-component (spatial) or per-(component, round) (spatio-temporal) \\
\bottomrule
\end{tabular}
\end{table*}

\subsubsection{Uniform Depolarizing Noise Model}
\label{subsec:unifom}
As a baseline, \sol{} provides a uniform depolarizing noise model~\cite{gidney2021fault} in which a single physical error parameter \(p\) is applied uniformly across all supported circuit locations. This includes single-qubit gates, two-qubit gates, reset operations, measurements, and idle intervals. The model introduces no operation-dependent, spatial, or temporal asymmetry and serves as a homogeneous reference against which more structured noise families can be compared.

\subsubsection{Pauli-Biased Noise Model}
\label{subsec:biased_xz}

Several experimentally relevant quantum-computing modalities exhibit dephasing-dominated noise~\cite{aliferis2009fault,chamberland2022circuit,tuckett2018ultrahigh,tuckett2019tailoring},  in which faults along one Pauli axis occur more frequently than along the orthogonal axis. Dephasing-dominated regimes, where phase-type faults dominate bit-flip faults, are the canonical case, but the reverse asymmetry arises in systems whose dominant channel is amplitude- or relaxation-driven. As a structured alternative to the homogeneous depolarizing baseline, \sol{} provides a biased $X$/$Z$ noise model, parameterized by a base error rate $p$, a bias axis $A\in\{X,Z\}$, and a bias factor $\eta = p_A / p_{A^\perp} \ge 1$ ($A^\perp$ denotes the orthogonal axis). This model is hardware-inspired and designed to isolate, along a single controlled axis, the effect of structured deviations from uniform depolarizing noise.

\paragraph{Channel forms.} For single-qubit and idle channels, the bias factor $\eta$ redistributes a fixed Pauli budget toward the dominant axis: $p_{A^\perp}=p_Y=p/(\eta+2)$ and $p_A=\eta p/(\eta+2)$. Two-qubit channels concentrate weight analogously on the three Paulis supported on $\{I,A\}$ and distinct from $II$, namely $\{AI, IA, AA\}$, with rates $\eta p/(12+3\eta)$ versus $p/(12+3\eta)$ on the remaining components. SPAM channels are basis-dependent: the bias-aligned and orthogonal flip probabilities are scaled to $p_{M_A}=2p/(1+\eta)$ and $p_{M_{A^\perp}}=2\eta p/(1+\eta)$, so that bias-aligned readout becomes high-fidelity as $\eta$ grows while orthogonal readout tracks the dominant channel. The total Pauli weight at every gate and idle location remains $p$, and the model reduces to uniform depolarizing noise at $\eta=1$.

\paragraph{Implications for Code Design.} Under sufficiently strong bias, symmetric codes become suboptimal. For CSS surface-code families, one may instead consider asymmetric code distances \(d_X\) and \(d_Z\), with $d_A > d_{A^\perp}$, to allocate more protection against the dominant $A$-type faults. More generally, bias-aware constructions such as the XZZX surface code~\cite{bonilla2021xzzx} achieve significantly improved thresholds and lower logical error rates in strongly biased regimes by tailoring the stabilizer checks to the underlying noise structure~\cite{tuckett2018ultrahigh, tuckett2019tailoring}.

\subsubsection{Measurement-Biased Noise Model}
\label{subsec:meas_biased_noise}

Physical error budgets are rarely distributed uniformly across operations, and measurement is typically slow and among the noisiest. For example, in superconducting surface-code experiments~\cite{google_below_threshold}, measurement and data-qubit idling during measurement are the largest contributors to the per-operation error budget, exceeding both single-qubit and CZ gate errors. In addition, measurement errors are particularly consequential because they corrupt the extracted syndrome, requiring repeated syndrome extraction to distinguish measurement faults from data errors; a common convention uses $d$ rounds per logical operation at distance $d$, e.g., following lattice surgery. To isolate this effect, \sol{} provides a measurement-biased family parameterized by a base error rate $p$ and a measurement-bias factor $\eta \ge 1$, targeting hardware regimes in which readout fidelity is the primary bottleneck.

\paragraph{Channel forms.} Single-qubit gates, two-qubit gates, and idling are held at the uniform-depolarizing baseline with rate $p$, while measurement and reset flip probabilities are rescaled to $p_{\mathrm{spam}}=\eta p$. The model reduces to uniform depolarizing noise at $\eta=1$.

\paragraph{Implications for syndrome extraction.} The measurement-biased model enables optimization of the SE round count under varying operational asymmetries~\cite{marton2025optimal}. For a given temporal window, the optimal round count is set by the balance between measurement, gate, and idling fidelities: measurement-dominated regimes benefit from additional rounds, since repeated readouts mitigate unreliable measurement outcomes, while gate- or idle-dominated regimes favor fewer rounds, since each additional round introduces operational faults that can outweigh the marginal utility of extra syndrome information. The model is also useful for evaluating QEC protocols that specifically target measurement-bias overhead~\cite{akahoshi2025runtime}.

\subsubsection{Non-Uniform Noise Model: Spatial and Temporal Variation}
\label{subsec:nonuniform_noise}

Realistic quantum processors do not generally exhibit spatially or temporally uniform noise, and the dominant mechanisms differ across platforms: fabrication imperfections, control inhomogeneity, and calibration drift in superconducting devices~\cite{tan2024resilience,spitz2018adaptive,mcewen2022resolving,auger2017fault}; zone-to-zone variation~\cite{hughes2025trapped} and control-parameter drift~\cite{gerster2022experimental} in trapped-ion processors; and site-to-site miscalibrations together with experimental imperfections, such as laser noise or inhomogeneities in neutral-atom arrays~\cite{evered2023high}.

To capture these effects, \sol{} provides a non-uniform noise family that models fluctuations across the device footprint and over the duration of an experiment. Both models are implemented by applying stochastic perturbations to a uniform baseline error scale \(p\).

\paragraph{Channel forms.} Every gate, SPAM, and idle channel of the uniform baseline (Section~\ref{subsec:builtin_noise}) is retained, but its rate $p$ is replaced by a per-component value $p_c = p(1+\delta_c)$ with $\delta_c \sim \mathcal{N}(0,\sigma^2)$, clipped to $[0,0.5]$ for physical validity. The index $c$ ranges over physical components for the spatial variant (sampled once per simulation) and over (component, round) pairs for the spatio-temporal variant (sampled per syndrome-extraction cycle). The model recovers uniform depolarizing noise at $\sigma=0$. Exact rate assignments are given in Table~\ref{tab:builtin_families}.

%% file: sections/Primitives.tex
\subsection{Surface Code Logical Primitives}

The second core component of \sol{} is a high-level circuit generator for surface code logical primitives. These primitives serve as the fundamental units of fault-tolerant computation and provide a basis for benchmarking how logical performance depends on code geometry, circuit scheduling, and hardware-specific noise. A primary metric for these evaluations is the spacetime volume ($V$), which quantifies the resource cost by taking the product of the spatial footprint (number of data qubits) and the temporal duration (number of syndrome-extraction rounds).

\sol{} automatically handles detector annotations and logical observable definitions. This allows for the seamless generation of Detector Error Models (DEMs) which can be passed directly to simulation and decoders pipeline. The current implementation supports four foundational primitives: logical memory, the transversal Hadamard, lattice surgery, and the lattice-surgery--based phase ($S$) gate.

\begin{figure}
    \centering
    \includegraphics[width=1\linewidth]{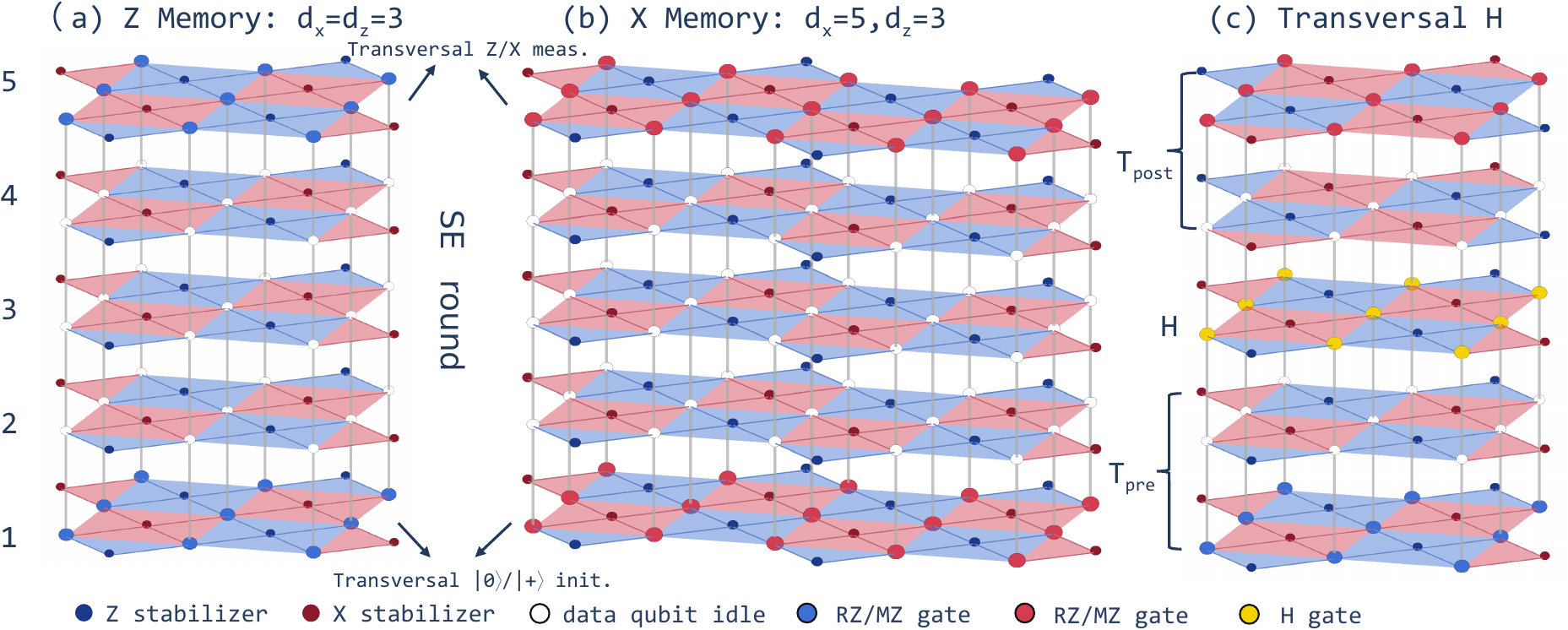}
\caption{\textbf{(a, b):} Spacetime structure of the logical memory primitive in the rotated surface code. A distance-($d_X, d_Z$) patch is transversally initialized in a logical eigenstate, undergoes $t$ rounds of stabilizer extraction (vertical axis), and is transversally measured in the same basis. Panel~(a) shows a $\bar{Z}$-basis experiment with $d_X = d_Z = 3$; panel~(b) shows its $\bar{X}$-basis counterpart. Blue and red data-qubit nodes mark transversal initialization and measurement in Z and X basis, respectively. \textbf{(c):} Logical Hadamard via transversal Hadamard on all data qubits, shown for a $\bar{Z} \to \bar{X}$ operation with $t_{\mathrm{pre}} = t_{\mathrm{post}} = 2$. Since the transversal Hadamard conjugates $X \leftrightarrow Z$, the $X$- and $Z$-type stabilizers swap after the gate, as indicated by the plaquettes color change.}

    \label{fig:memory-primitive}
\end{figure}

\subsubsection{Logical Memory}
\label{subsec:memory}
The memory experiment benchmarks the ability of a logical qubit to preserve quantum information while stationary, and serves as the baseline for estimating the logical error rate (LER). As shown in Fig.~\ref{fig:memory-primitive}(a,b), a $Z$- or $X$-basis memory experiment transversally initializes data qubits of a rotated surface-code patch in eigenstate ($|0\rangle_L$ or $|+\rangle_L$), performs $t$ rounds of syndrome extraction, and measures the patch transversally in the same basis. Configurable parameters include:
\begin{itemize}
    \item \textbf{Distances ($d_X, d_Z$):} The code distances along each boundary, enabling protection against phase and bit-flip chains to be tuned independently.
    \item \textbf{SE Rounds ($t$):} The number of repeated syndrome-extraction cycles.
    \item \textbf{Basis:} The initialization and measurement axis ($\bar{X}$ or $\bar{Z}$).
\end{itemize}
The spacetime volume for logical memory is defined simply as: $V_{\mathrm{mem}} = d_X d_Z t$.

\subsubsection{Transversal Hadamard Gate}
\label{subsec:h}
The logical Hadamard ($H_L$) primitive implements a transversal operation, a native capability of the rotated surface code. \sol{} constructs this primitive by applying physical Hadamard gates to every data qubit within the logical patch, as depicted in Fig.~\ref{fig:memory-primitive}(c). This operation maps $\bar{X} \leftrightarrow \bar{Z}$ while interchanging the roles of $X$- and $Z$-type stabilizers. The generator automatically tracks this transformation, ensuring that subsequent rounds reflect the rotated patch orientation and swapped parity checks. Parameters include:
\begin{itemize}
    \item \textbf{Patch distances ($d_X, d_Z$):} Patch dimensions, allowing for asymmetric protection before and after the gate.
    \item \textbf{Pre-gate SE rounds ($t_{\mathrm{pre}}$):} repeated cycles performed before the transversal layer to establish the input state.
    \item \textbf{Post-gate SE rounds ($t_{\mathrm{post}}$):} repeated cycles performed after the transversal layer to verify logical integrity.
\end{itemize}
The spacetime volume mirrors the memory experiment: $  V_{\mathrm{Had}} = d_X d_Z (t_{\mathrm{pre}} + t_{\mathrm{post}} + 1)$.

\subsubsection{Lattice Surgery}
\label{subsec:lattice surgery}
Lattice surgery is the primary mechanism for realizing entangling operations through joint logical parity measurements ($M_{XX}$ or $M_{ZZ}$). 
\sol{} automates the construction of the resulting 3D spacetime structure, illustrated in Fig.~\ref{fig:ls-primitive}(a,b). Pre-surgery, the two patches are stabilized by independent generator sets $\mathcal{S}_1$ and $\mathcal{S}_2$. During the merge phase, the bridge data qubits are initialized in the basis orthogonal to the surgery basis (blue/red nodes in Fig.~\ref{fig:ls-primitive} for $M_{XX}$ / $M_{ZZ}$ surgery, respectively), and the stabilizer set is expanded to $\mathcal{S}_{\mathrm{merge}}$ by introducing new weight-2 and weight-4 stabilizer along an ancilla bridge of length $L$; their product yields the joint logical parity $\bar{Z}_1 \bar{Z}_2$ (or $\bar{X}_1 \bar{X}_2$), which is extracted over $t_{\mathrm{merge}}$ syndrome rounds. The split phase measures the bridge data qubits in the same basis used for their initialization, contracting the stabilizer set back to $\mathcal{S}_1 \cup \mathcal{S}_2$ for $t_{\mathrm{post}}$ post-surgery rounds and yielding the characteristic ``H-shaped'' spacetime topology. Configurable parameters include:
\begin{itemize}
    \item \textbf{Patch distances ($d_X, d_Z$):} The dimensions of the participating logical patches.
    \item \textbf{Ancilla path length ($L$):} The spatial distance between the logical patches, representing routing overhead.
    \item \textbf{Temporal rounds ($t_{\mathrm{pre}}, t_{\mathrm{merge}}, t_{\mathrm{post}}$):} Rounds performed for each phase. $t_{\mathrm{merge}}$ provides crucial redundancy against time-like errors during the joint measurement.
\end{itemize}
The total spacetime volume $V_{\mathrm{LS}}$ is the sum of its constituent phases:
$V_{\mathrm{LS}} = 2 d_X d_Z (t_{\mathrm{pre}} + t_{\mathrm{post}}) + (2 d_X d_Z + L \cdot d_{b}) t_{\text{merge}}, \text{ with } d_{b} = d_X \text{ for } M_{ZZ} \text{ and } d_Z \text{ for } M_{XX}.
$

\subsubsection{Logical Phase Gate}
\label{subsec:S}
\begin{figure}
    \centering
    \includegraphics[width=1\linewidth]{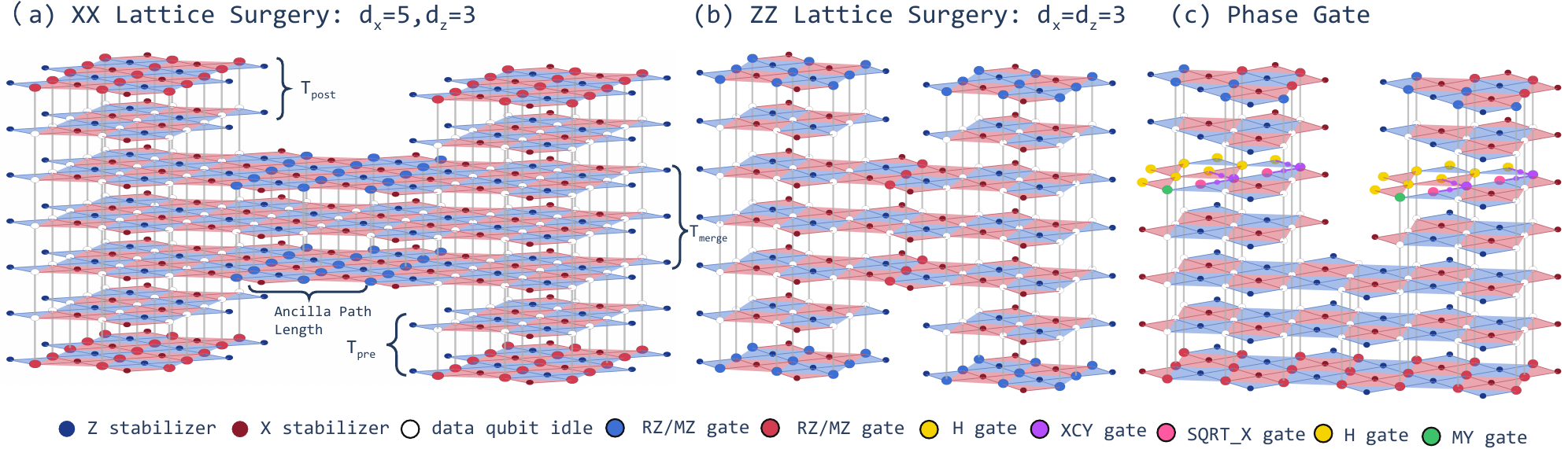}
    \caption{\textbf{(a) and (b):} Spacetime structure of the lattice-surgery primitive. Two logical patches (pre-surgery columns) are temporarily merged for $T_{\mathrm{merge}}$ rounds along an ancilla bridge of length $L$ to extract the joint logical parity ($M_{XX}$ or $M_{ZZ}$), then split apart for $T_{\mathrm{post}}$ post-surgery rounds. The merge--split bridge yields the characteristic ``H-shaped'' spacetime topology. \textbf{(c)} Logical phase-gate: one logical data patch and one logical ancilla patch are initialized in the logical $X$ basis, after which a logical $M_{ZZ}$ lattice-surgery measurement is performed between them. The ancilla patch is then measured in the logical $Y$ basis using the $Y$ transition round from Ref.~\cite{Gidney2024inplaceaccessto} followed by additional boundary rounds that suppress measurement errors introduced during the transition. Conditioned on the lattice-surgery and ancilla-measurement outcomes, this protocol implements a logical $S$ gate on the data patch. Since $S$ maps an $X$-basis state to a $Y$-basis state, we subsequently measure the data patch in the logical $Y$ basis.}
    \label{fig:ls-primitive}
\end{figure}

The logical phase gate ($S_L$) is non-transversal in the rotated surface code and requires a measurement-based construction. \sol{} implements $S_L$ using the in-place $Y$-basis measurement protocol of Ref.~\cite{Gidney2024inplaceaccessto}, illustrated in Fig.~\ref{fig:ls-primitive}(c). The sequence is: (1)~the data patch and an adjacent ancilla patch are both initialized in the logical $\bar{X}$ basis; (2)~a logical $M_{ZZ}$ parity measurement is performed between them via lattice surgery, transiently expanding the stabilizer set as in \S\ref{subsec:lattice surgery}; (3)~the ancilla is measured in the logical $\bar{Y}$ basis using the $Y$ transition round of Ref.~\cite{Gidney2024inplaceaccessto}, followed by $t_{\mathrm{boundary}}$ boundary rounds that suppress measurement errors introduced during the transition. Conditioned on the surgery and ancilla outcomes, this realizes $S_L$ on the data patch; since $S$ maps an $\bar{X}$ eigenstate to a $\bar{Y}$ eigenstate, the data patch is subsequently read out in the logical $\bar{Y}$ basis. To isolate the intrinsic cost of the gate logic, the data-patch $\bar{Y}$ readout is performed as a noiseless \texttt{MPP} of the logical $Y$ operator.

For this primitive, the patches are constrained to be square ($d_X = d_Z = d$). Configurable parameters include:
\begin{itemize}
    \item \textbf{Code distance ($d$):} The side length of the square logical patches.
    \item \textbf{Ancilla path length ($L$):} The spatial separation between the data and ancilla patches.
    \item \textbf{Merge rounds ($t_{\mathrm{merge}}$):} SE rounds during the lattice-surgery $M_{ZZ}$ step.
    \item \textbf{Padding rounds ($t_{\mathrm{boundary}}$):} Extra rounds performed after the $\bar{Y}$ transition to protect against time-like errors.
\end{itemize}
The spacetime volume is dominated by the two-patch surgery and the padded ancilla readout: $V_{\mathrm{Phase}} = (2 d^2 + L \cdot d)\, t_{\mathrm{merge}} + d^2\, t_{\mathrm{boundary}}$.

All primitives are instantiated from user-specified parameters, emitted as \textsc{Stim} circuits with detector annotations and logical observables, and consumed directly by the simulation pipeline of \S\ref{sec:evaluation} to estimate logical thresholds and error suppression under the structured noise families of \S\ref{subsec:builtin_noise}.

Together, these primitives form a foundational set of operations for evaluating logical performance, resource scaling, and noise sensitivity. All primitives can be instantiated with user-specified parameters, exported as high-performance \textsc{Stim} circuits with detector annotations and logical observables, and consumed directly by the simulation pipeline of \S\ref{sec:evaluation} to estimate logical thresholds and error suppression under the structured noise families of \S\ref{subsec:builtin_noise}.

%% file: sections/evaluation.tex
\section{Evaluation}
\label{sec:evaluation}
We evaluate the logical primitives generated by \sol{} using Monte Carlo stabilizer simulation with \textsc{Stim}~\cite{gidney2021stim}, sweeping the code distance $d\in\{3,5,7,9,11\}$. Each configuration is sampled until either $10^7$ shots or $10^3$ logical errors are accumulated with the shot cap raised to $10^8$ for low-LER points (typically $d=11$ at small $p$). Syndrome data is decoded with minimum-weight perfect matching via \textsc{PyMatching}~\cite{higgott2022pymatching}, using the uncorrelated variant unless stated otherwise. 

We report two metrics: (i)~the absolute logical error rate per round (LER$/$round), and (ii)~the relative LER, defined as the ratio between the LER under a structured noise model and the LER under the uniform-depolarizing baseline at the same physical error rate $p$. As described in Section~\ref{sec:primitivebench}, the baseline assigns the same $p$ to SPAM, gate error and idling errors. A structured noise model is from one of the three hardware-motivated families: \emph{Pauli $Z$-biased noise}, in which every depolarizing channel is replaced by a homogeneous asymmetric Pauli channel with weights $p_X{=}p_Y{=}p/(\eta{+}2)$, $p_Z{=}\eta p/(\eta{+}2)$ and bias factor $\eta\!\ge\!1$; \emph{measurement-biased noise}, in which only the per-measurement Pauli-flip rate is rescaled to $\eta p$---gate, reset, and idling channels held at the baseline---homogeneously across qubits and rounds with $\eta\!\ge\!1$; and \emph{non-uniform depolarizing noise},  in which each component receives an independently drawn rate $p_i = p\,(1+\sigma\,\xi_i)$ with $\xi_i\!\sim\!\mathcal{N}(0,1)$, sampled once per qubit (``space-only'') or once per (qubit, round) pair (``spatio-temporal''), with heterogeneity strength $\sigma\!\ge\!0$. We apply each family in turn to logical memory, the logical Hadamard, lattice surgery, and the logical phase gate, to highlight how the same underlying noise structure produces qualitatively different logical behavior across primitives.

\subsection{Logical Memory}

Logical memory, introduced in \S\ref{subsec:memory}, provides the baseline primitive for characterizing how a surface-code patch preserves an encoded state across repeated syndrome-extraction rounds: a logical eigenstate is initialized, protected for $t$ rounds, and measured in the same basis.

\subsubsection{Z-biased noise}

Here, we consider both square patches and rectangular patches with $d_Z > d_X$, allowing protection against $Z$-type and $X$-type logical fault chains to be tuned independently. The number of syndrome-extraction rounds is matched to the distance protecting the basis under test: $t = d_X = d_Z$ for square patches, and for rectangular patches $t = d_Z$ in $X$-basis memory and $t = d_X$ in $Z$-basis memory.

\begin{figure}
    \centering
    \includegraphics[width=0.99\linewidth]{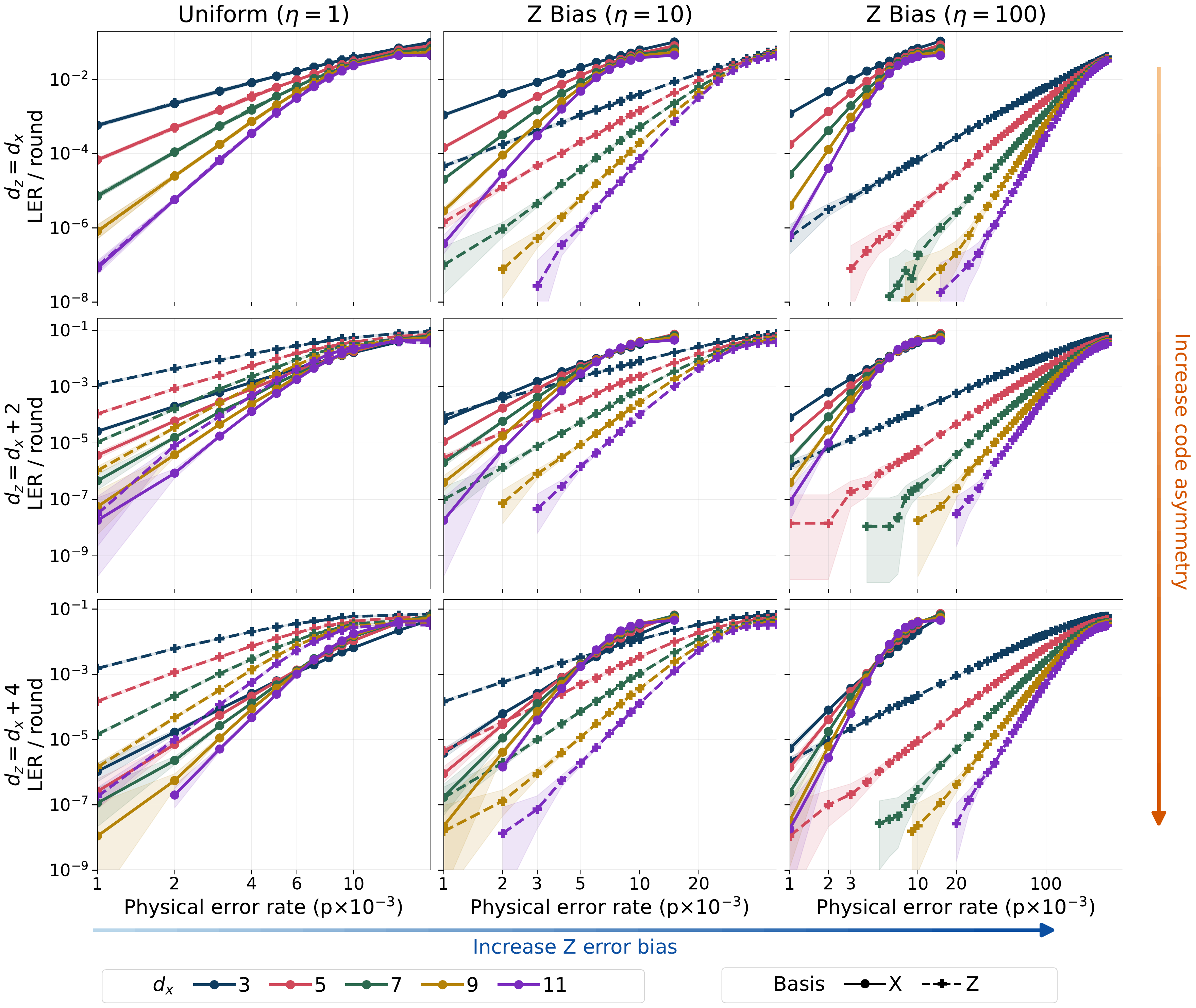}
    \caption{Logical memory under uniform depolarizing ($\eta=1$) and $Z$-biased noise ($\eta\in\{10,100\}$). Each panel plots LER per round versus physical error rate for both bases; columns sweep $\eta$ and rows increase patch asymmetry $d_Z-d_X\in\{0,2,4\}$. The $X$-basis experiment, limited by $Z$-driven fault chains, is suppressed by enlarging $d_Z$. The $Z$-basis experiment, limited by $d_X$, improves error suppression as $p_X=p_Y\propto 1/\eta$ shrinks the dominant fault channel.}

    \label{fig:memory-z-biased}
\end{figure}

Figure~\ref{fig:memory-z-biased} shows logical memory under uniform depolarizing noise and increasingly strong Pauli $Z$ bias. Under uniform depolarizing noise, square patches exhibit the expected symmetry between logical $X$- and $Z$-basis memory. When the noise becomes $Z$ biased, this symmetry is broken. The physical $Z$ error rate increases from $p/3$ in the uniform case to $p_Z = \frac{\eta}{\eta+2}p$, which remains $O(p)$ but is enhanced by a constant factor. For example, $p_Z = \tfrac{10}{12}p$ for $\eta=10$ and $p_Z = \tfrac{100}{102}p$ for $\eta=100$. In contrast, $X$- and $Y$-type errors are suppressed as $ p_X = p_Y = \frac{1}{\eta+2}p = O\!\left(\frac{p}{\eta}\right)$.

Physical $Z$ faults are more likely, and chains of such faults implement logical $Z$ errors, which flip an encoded $X$-basis state. Consequently, logical $X$-basis memory becomes more sensitive to the dominant noise channel as the bias grows. Increasing $d_Z$ directly suppresses these dominant $Z$-driven fault chains by increasing the minimal weight required to cause a logical error, leading to a substantial reduction in LER for $X$-basis memory under strong bias. At the same time, the logical $Z$-basis experiment benefits from the suppression of physical $X$- and $Y$-type errors, but remains primarily limited by the shorter distance $d_X$ in rectangular patches.

As a result, the apparent threshold behavior of the $X$- and $Z$-basis experiments can differ substantially under biased noise. Changing the aspect ratio $(d_X,d_Z)$ can vary protection between the two logical channels, consistent with recent hardware experiments~\cite{vezvaee2025surface}, but it does not fully restore the symmetry present under uniform depolarizing noise. This illustrates that patch geometry becomes a key architectural parameter under biased noise: a geometry that suppresses the dominant fault channel in one logical basis may provide weaker protection for the conjugate logical basis.

\subsubsection{Decoder Comparison}

\begin{figure}
    \centering
    \includegraphics[width=0.98\linewidth]{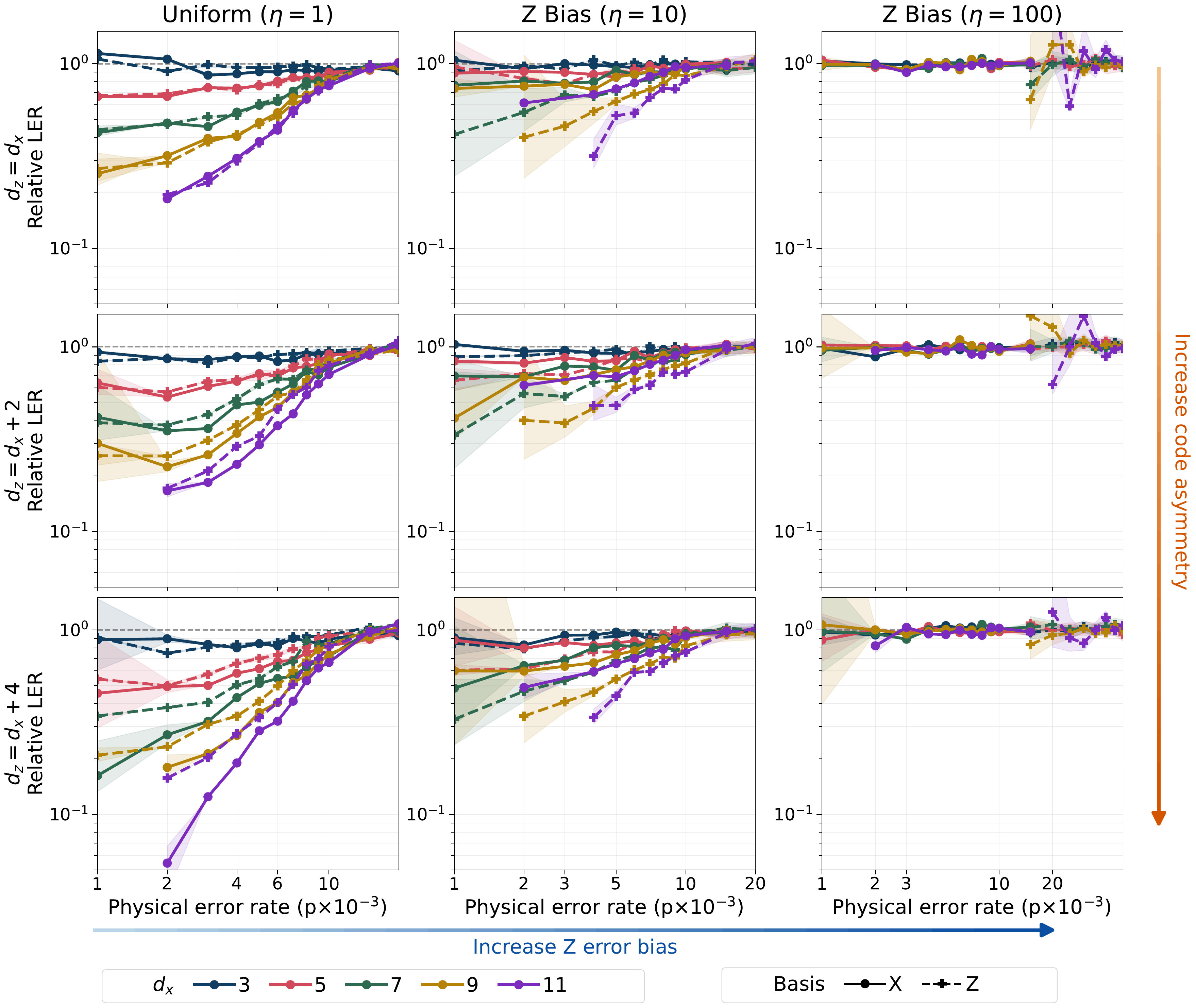}
    \caption{Relative Logical Error Rate (LER) for memory experiments comparing correlated PyMatching \cite{fowler2013optimal} to uncorrelated PyMatching \cite{higgott2022pymatching} under a $Z$-biased noise model ($\eta = 1$ denotes uniform depolarizing noise). Results are shown for varying code geometries. It is observed that the relative advantage of correlated PyMatching decreases as the noise becomes increasingly biased.}
    \label{fig:corre-pymatching}
\end{figure}

Figure~\ref{fig:corre-pymatching} fixes the noise family at Pauli $Z$-biased noise and varies the decoder, comparing correlated matching~\cite{fowler2013optimal} against the uncorrelated PyMatching~\cite{higgott2025sparse} default used elsewhere in this section. The bias factor $\eta\in\{1,10,100\}$ is swept on the same square and rectangular memory patches as above, and the relative LER between the two decoders is reported at matched configuration $(d_X,d_Z,p,\eta)$.

It's shown that correlated matching consistently yields a lower LER than uncorrelated matching. Uncorrelated matching decodes the $X$- and $Z$-type syndrome graphs independently, so a single physical $Y$ fault, which flips both an $X$-type and a $Z$-type detector, is treated as two unrelated errors and incurs twice the matching weight it deserves. Correlated matching instead couples the two graphs: after producing a tentative correction on one graph, it reweights edges on the other to reflect the conditional probability that a fault already matched as $X$ (or $Z$) was in fact a $Y$. This recovers the $X{\oplus}Z$ correlation structure of the depolarizing channel, lowering the effective weight of $Y$-induced detection events and the resulting logical error rate. 

The advantage is largest under uniform depolarizing noise ($\eta=1$), where $p_Y=p_X=p_Z$ and the off-diagonal correlations carry their full weight. The relative LER decreases with increasing code distance and decreasing $p$, where logical failures are dominated by rare high-weight error chains. For square patches ($d_X=d_Z$) the relative-LER curves for $X$- and $Z$-basis memory match, consistent with the basis symmetry of the underlying Pauli channel. For rectangular patches ($d_Z>d_X$) under uniform depolarizing noise, the two bases separate at low $p$: the $X$- and $Z$-basis experiments are protected by different code distances, so the high-weight error chains that dominate at low $p$ scale differently with $p$ in each basis, and the correlated-matching benefit tracks that asymmetry.

As $\eta$ grows, $p_Y\propto 1/\eta$ shrinks and the off-diagonal correlations exploited by correlated matching diminish. By $\eta=10$ the $X$-basis relative LER has already risen compared to the relative LER at the uniform-depolarizing noise, indicating a smaller correlated-matching advantage. At $\eta=100$, the $X$-basis experiment is swept over $p\in[10^{-3},10^{-2}]$ and the $Z$-basis experiment over $p\in[10^{-2},5\times10^{-2}]$; the $Z$-basis absolute LER falls below $\sim\!10^{-8}$ for $p<10^{-2}$, where accurate estimation would require prohibitively many shots. Across the swept ranges the two decoders deliver similar LER, though the gap may widen at lower $p$ that is currently inaccessible at this sampling budget. This suggests decoder performance is itself shaped by the underlying error structure: as the channel becomes more biased and off-diagonal mass shrinks, the correlated-matching advantage narrows and the cheaper uncorrelated decoder becomes increasingly competitive.

\subsubsection{Measurement-biased noise}
\begin{figure}[t]
    \centering

    \begin{subfigure}{0.99\linewidth}
        \centering
        \includegraphics[width=\linewidth]{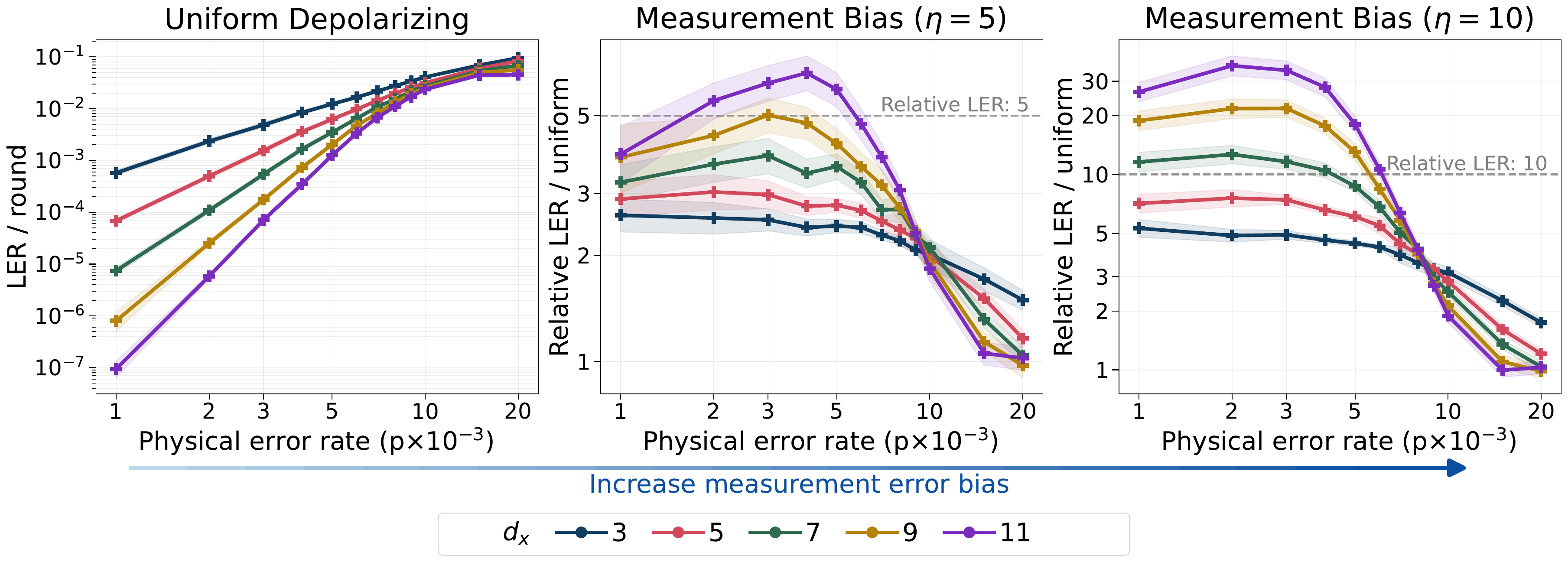}
        \caption{Varying physical error rate at rounds $=d$. Relative overhead peaks near threshold and is amplified at larger distances.}
        \label{fig:memory-meas-relative-phys}
    \end{subfigure}

    \vspace{0.8em}

    \begin{subfigure}{0.99\linewidth}
        \centering
        \includegraphics[width=\linewidth]{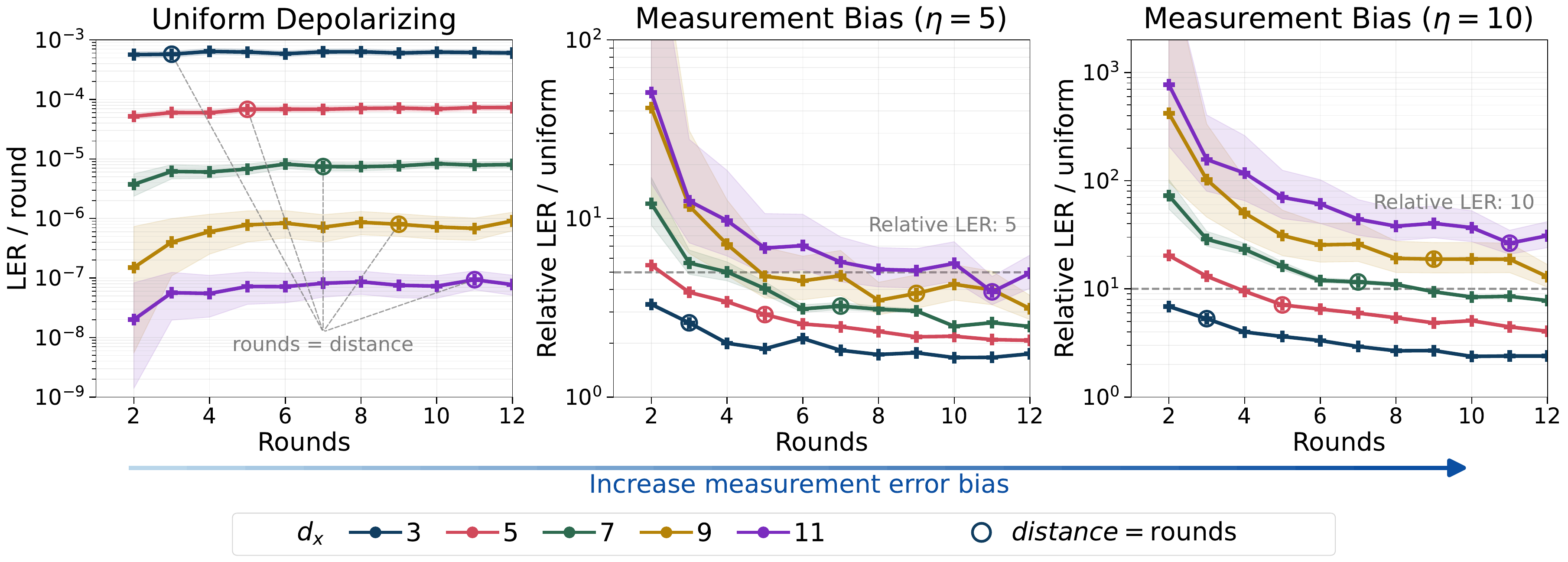}
        \caption{Varying rounds at fixed $p=10^{-3}$. Boundary-limited behavior at small round count relaxes to the bulk as rounds approach and exceed $d$.}
        \label{fig:memory-meas-relative-rounds}
    \end{subfigure}

    \caption{Logical memory under measurement-biased noise, normalized to the uniform-depolarizing baseline. In each subfigure the left column shows the absolute baseline LER per round, while the middle and right columns show relative LER for $\eta\in\{5,10\}$ on square patches. Relative overhead grows with the measurement bias, is largest near the threshold regime, and is recovered by adding syndrome-extraction rounds. Measurement-dominated hardware therefore benefits from temporal-redundancy headroom that is invisible under uniform-depolarizing analyses.}
    \label{fig:memory-meas-stacked}
\end{figure}

Figure~\ref{fig:memory-meas-stacked} evaluates logical memory under measurement-biased noise from two complementary views: varying the physical error rate at fixed $t=d$ (Fig.~\ref{fig:memory-meas-relative-phys}), and varying the number of syndrome-extraction rounds at fixed $p=0.001$ (Fig.~\ref{fig:memory-meas-relative-rounds}). 

In Fig.~\ref{fig:memory-meas-relative-phys}, the number of rounds is fixed to $T=d$. Therefore, increasing the measurement-bias factor increases the logical error rate relative to the uniform-depolarizing baseline. Near threshold, the relative LER is suppressed, reflecting the overall degradation of code performance near threshold, where error correction is no longer effective. At low physical error rates, we observe a pronounced effect of measurement bias, particularly for higher code distances. At this regime, logical errors arise from long, rare chains. Measurement bias skews these chains over time, and this effect accumulates more for higher-distance codes, making them more sensitive to the bias.


Figure~\ref{fig:memory-meas-relative-rounds} shows the logical error rate per round as a function of the number of rounds at fixed physical error rate $p=0.001$. The left panel corresponds to the absolute value for uniform depolarizing noise model. For the uniform baseline, the LER increases slightly with the number of rounds, especially at larger code distances, before saturating. This is a boundary effect~\cite{qcse_rounds_distance}: near spatial and temporal boundaries(initialization and measurement), error chains are truncated, reducing their number compared to the bulk. As the code distance increases, longer error chains dominate and require the initialization and measurement boundaries to be placed farther apart, i.e., a larger number of rounds is needed. As a result, finite-round experiments are boundary-limited, while the relevant threshold is the bulk value. This is consistent with Fig.~2 of~\cite{google2023suppressing}. In addition, we annotate the case where the number of rounds equals the code distance, which moves the system away from boundary effects. We observe that measurement bias has a stronger impact when the rounds are too few, while the relative LER saturates once enough rounds are used. This highlights the importance of using enough rounds to reach the bulk regime.

\subsubsection{Non-uniform noise}

\label{subsec:eval_memory_nonuniform}

\begin{figure}
    \centering
    \includegraphics[width=0.99\linewidth]{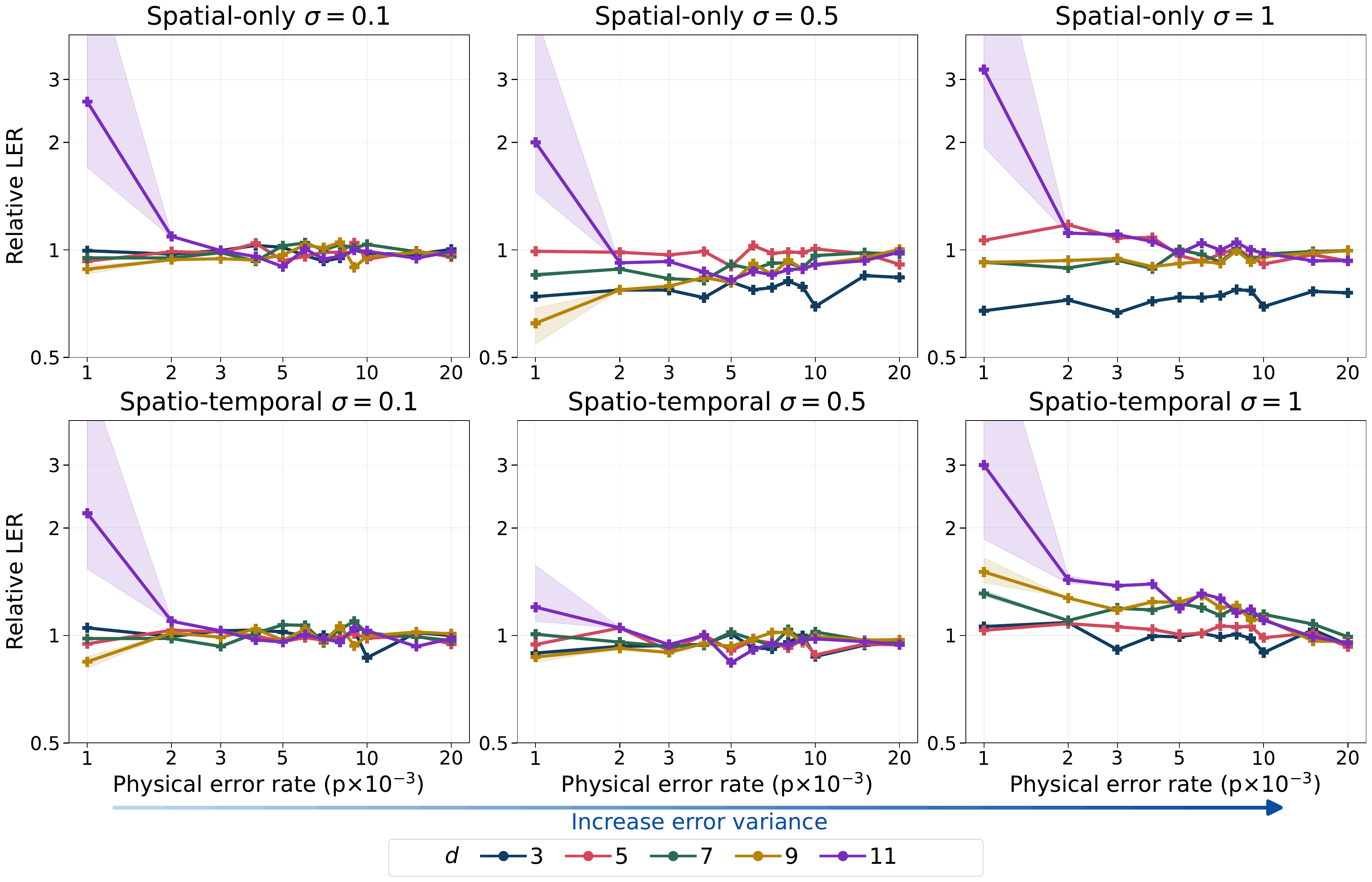}
    \caption{Logical memory under non-uniform noise, normalized to the uniform-depolarizing baseline on square patches in the $Z$-basis. Heterogeneity strength $\sigma$ scales a Gaussian perturbation applied to the per-component error rate.}
    \label{fig:memory-nonuniform-combined}
\end{figure}

The non-uniform experiments probe how logical memory responds when the physical error rate varies across space only or across both space and time. Figure~\ref{fig:memory-nonuniform-combined} reports the logical error rate relative to the uniform-depolarizing baseline (left panel of Fig.~\ref{fig:memory-meas-relative-phys}); non-uniformity is modeled by sampling a per-component Gaussian perturbation factor $\gamma$ applied to $p$ as introduced in \S\ref{subsec:nonuniform_noise}, and the decoder prior is matched to the realized per-component rates.

Across the surveyed regime, relative LER stays close to the uniform-depolarizing baseline. The curves at $\gamma\in\{0.1,0.5,1\}$ are largely overlapping even at the largest distances plotted, indicating that, with matched decoder priors, logical memory absorbs moderate spatial heterogeneity without a systematic LER penalty. Spatial-only and spatio-temporal disorder track each other closely, so adding round-to-round variation on top of spatial heterogeneity does not introduce a qualitatively new failure mode at the variances studied here. 

A subset of curves dip below relative LER $=1$ for $\sigma = 1$, particularly at smaller code distances. For each data point, a single draw of per-component perturbations is used to instantiate one noisy circuit used to sample for LER. Smaller patches contain fewer physical components, so the realized sample mean of $p_c$ has higher variance from one draw to the next and may fall slightly above or below the nominal $p$. This sampling stochasticity likely contributes to the sub-unity relative LER seen at smaller distances; averaging over many independent perturbation draws would be expected to smooth out the fluctuation.

\subsection{Hadamard}
The transversal Hadamard, introduced in \S\ref{subsec:h}, exchanges the logical $\bar{X}$ and $\bar{Z}$ operators by applying physical Hadamards to every data qubit, swapping the roles of $X$- and $Z$-type stabilizers; we benchmark it as a transversal layer with $t_{\mathrm{pre}}$ and $t_{\mathrm{post}}$ syndrome-extraction rounds.
\subsubsection{Z-biased noise}
\begin{figure}
    \centering
    \includegraphics[width=0.99\linewidth]{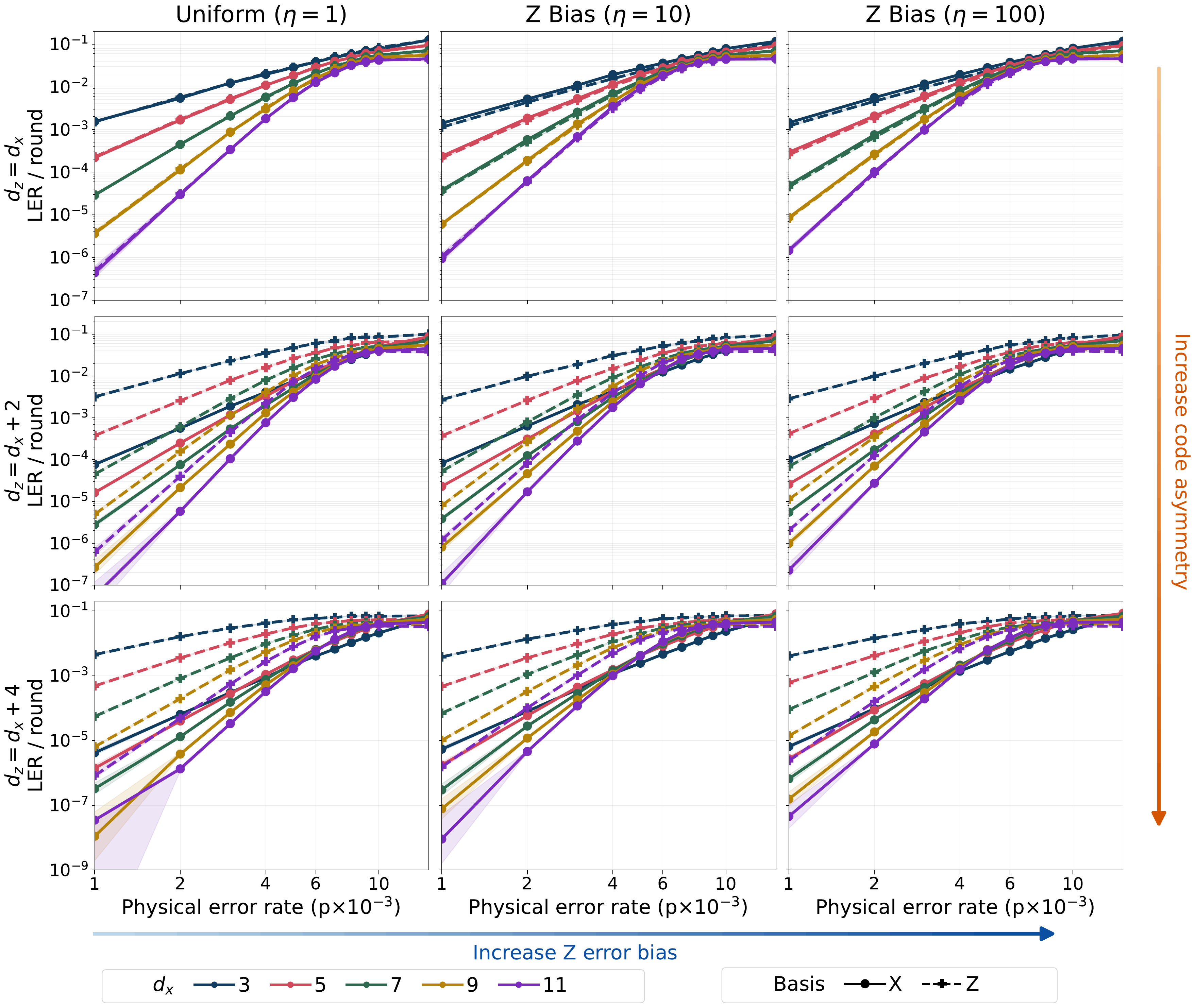}
    \caption{Logical Hadamard under uniform depolarizing and $Z$-biased noise. Each panel plots LER per round versus physical error rate for both logical bases; columns sweep $\eta\in\{1,10,100\}$ and rows increase the patch asymmetry by growing $d_Z$. Unlike memory, the two basis curves remain nearly symmetric even at $\eta=100$, and the benefit of $d_Z>d_X$ is strongly diminished. The transversal Hadamard exchanges $X$ and $Z$ channels mid-circuit and effectively averages the bias across the two axes, dissolving the asymmetry that geometry-tailored distances otherwise exploit.}
    \label{fig:h-z-biased}
\end{figure}

Figure~\ref{fig:h-z-biased} shows the logical Hadamard, where the number of rounds before and after the transversal Hadamard is set equal to the code distance. Its response to $Z$-biased noise is qualitatively different from that of logical memory.

The key reason is that the Hadamard layer swaps $X$ and $Z$ errors midway through the circuit, so the primitive does not preserve the input bias. In other words, a $Z$-biased noise channel before the Hadamard becomes effectively $X$-biased afterward. As a result, the circuit mixes the error basis rather than maintaining a consistent asymmetry, making the Hadamard inherently non--bias-preserving.

Consequently, we observe nearly symmetric behavior between $X$ and $Z$ experiments, even under strongly $Z$-biased noise. While patch asymmetry still has an effect, the advantage of increasing $d_Z$ is partially washed out, since the bias is redistributed across both error types during the operation. This reduces the extent to which the code can exploit the underlying noise asymmetry.

This makes the Hadamard a useful counterexample: the benefit of biased noise is not determined by code geometry alone, but also by the structure of the logical operation. Primitives that preserve the error basis (such as memory) can directly leverage bias, while those that mix bases (such as Hadamard) can significantly diminish or even neutralize that advantage.

\subsubsection{Measurement-biased noise}
\begin{figure}
    \centering

    \begin{subfigure}[t]{\linewidth}
        \centering
        \includegraphics[width=0.99\linewidth]{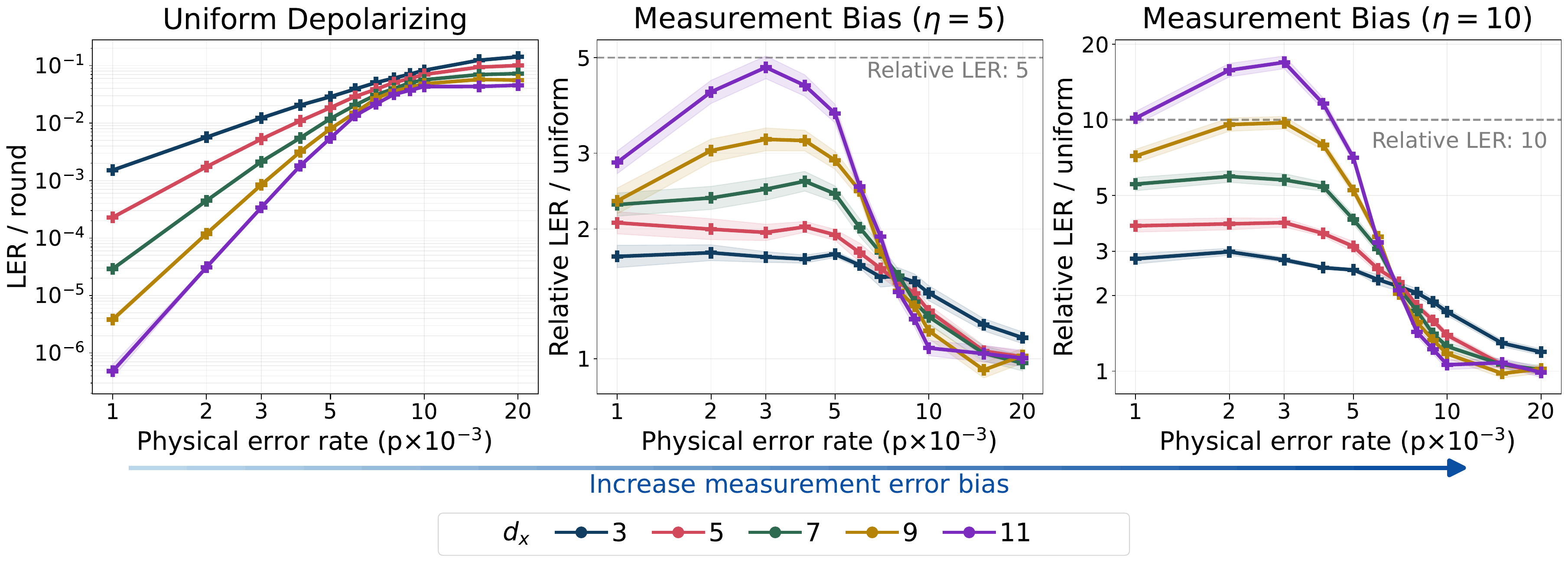}
        \caption{Varying physical error rate at rounds $=d$.}
        \label{fig:h-meas-relative-p}
    \end{subfigure}

    \vspace{0.5em}

    \begin{subfigure}[t]{\linewidth}
        \centering
        \includegraphics[width=0.99\linewidth]{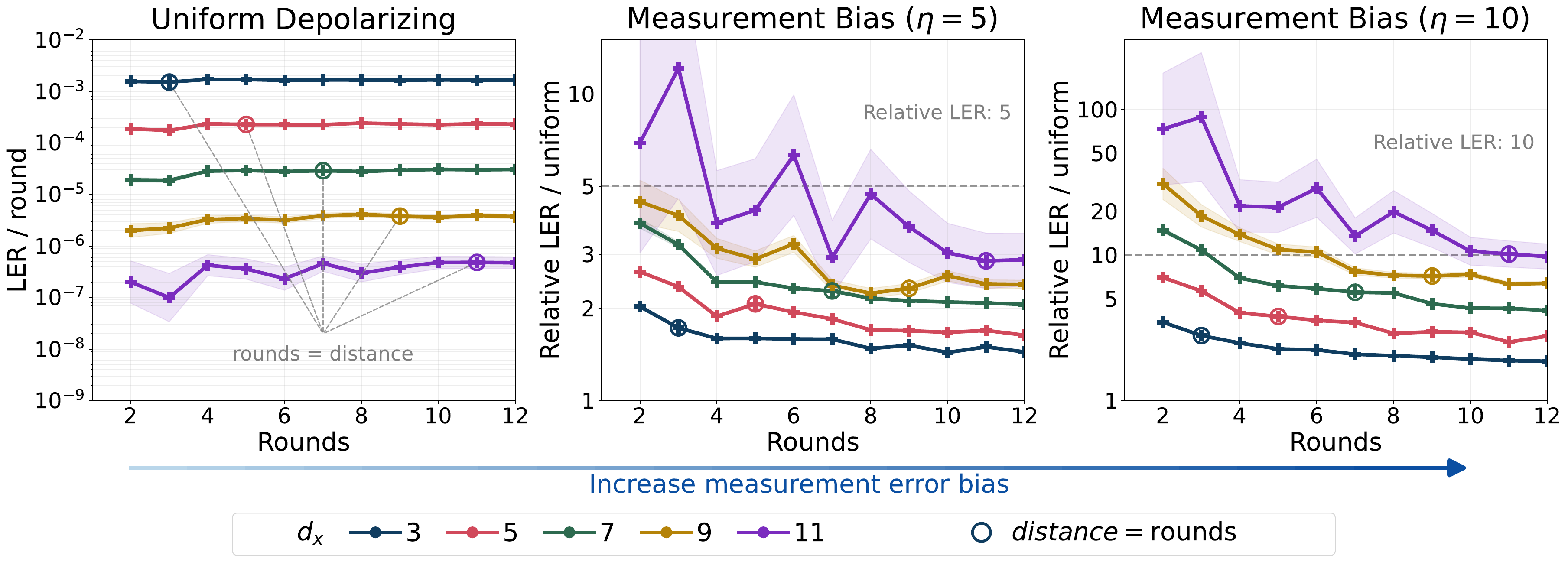}
        \caption{Varying rounds at fixed $p$. The strongest bias remains furthest above the baseline and relaxes most slowly with added rounds.}
        \label{fig:h-meas-relative-rounds}
    \end{subfigure}

    \caption{Logical Hadamard under measurement-biased noise, normalized to the uniform-depolarizing baseline on square patches. The left column of each subfigure shows the absolute baseline LER per round, while the middle and right columns show relative LER for $\eta\in\{5,10\}$. Relative overhead grows with the bias, peaks at intermediate physical error rates, and is recovered only slowly by adding rounds. The transversal layer keeps Hadamard's temporal-redundancy requirement comparable to memory's, but shifts the peak-penalty regime closer to threshold.}
    \label{fig:h-meas-relative-combined}
\end{figure}

Figure~\ref{fig:h-meas-relative-combined} shows the absolute logical error rate (LER) for the uniform depolarizing baseline and the relative LER under measurement-biased noise with bias factors $\eta=5$ and $\eta=10$ for the Hadamard primitive. Figure~\ref{fig:h-meas-relative-p} demonstrates that measurement bias increases the logical error rate relative to the uniform baseline across the full range of physical error rates.

Consistent with the memory experiment, for a fixed number of rounds, the relative LER depends non-monotonically on the physical error rate. In particular, the largest degradation occurs at intermediate physical error rates (around $3\times10^{-2}$ for $\eta=5,10$), while the relative LER is reduced at lower $p$. This can be understood as an interplay between suppression and proximity to threshold: although measurement errors are correctable, their impact is maximized when the effective error rate approaches the threshold, where the code is most vulnerable.

Figure~\ref{fig:h-meas-relative-rounds} shows how this overhead evolves with the number of rounds. As in the memory case, stronger measurement bias leads to a larger logical penalty, and increasing the number of rounds reduces this effect until saturation. When the number of rounds is too small, the system remains boundary-limited and the impact of measurement bias is more pronounced, while additional rounds move the system closer to the bulk regime.

\subsubsection{Non-uniform noise}
\label{subsec:eval_h_nonuniform}
\begin{figure}
    \centering
    \includegraphics[width=0.9\linewidth]{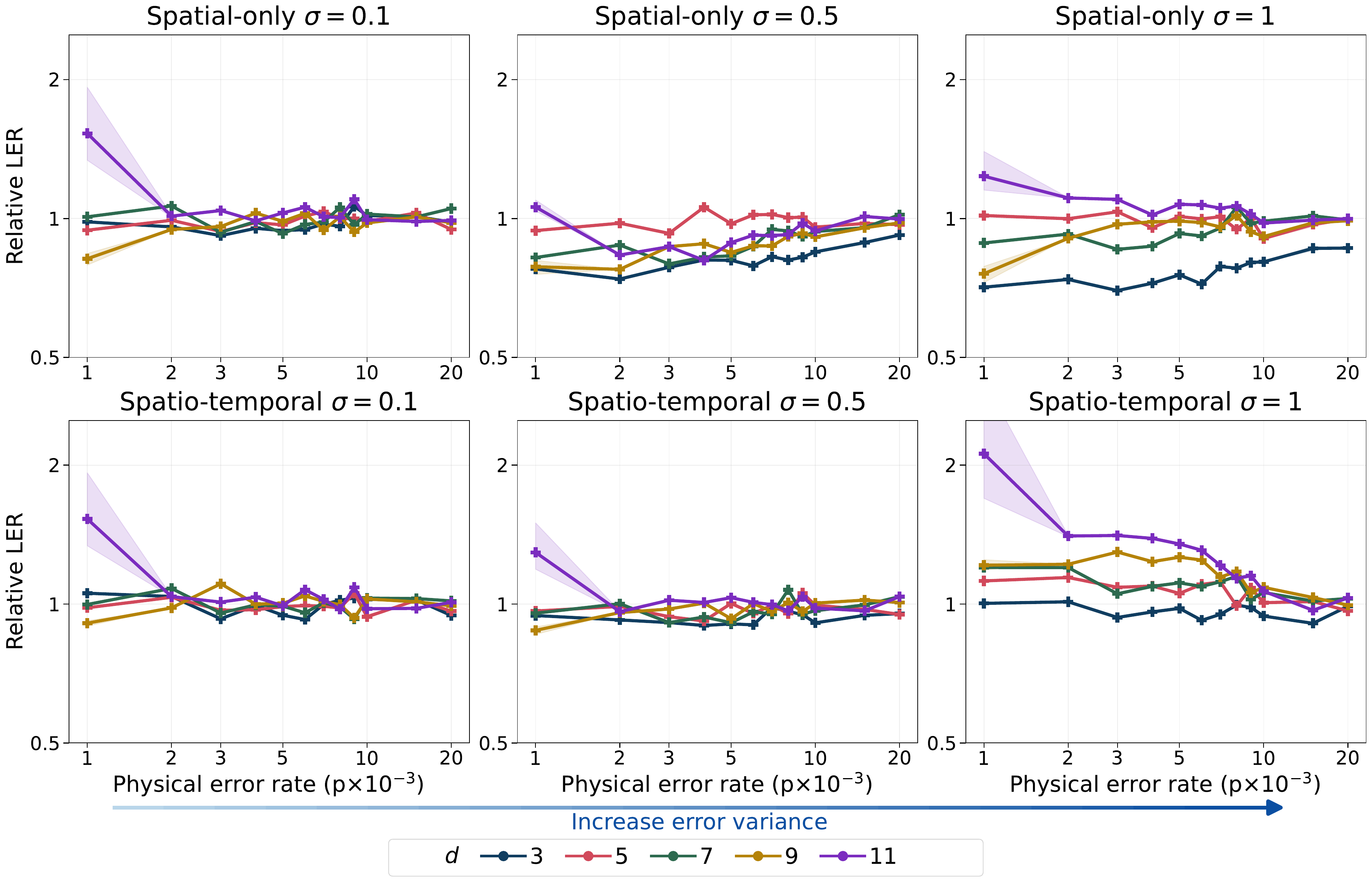}
    \caption{Logical Hadamard under non-uniform noise, normalized to the uniform-depolarizing baseline on square patches in the $Z$-basis. }
    \label{fig:h-nonuniform}
\end{figure}

Figure~\ref{fig:h-nonuniform} examines the transversal Hadamard's response to noise heterogeneity. The aggregate behavior matches memory: relative LER stays within roughly a factor of two of the uniform-depolarizing baseline across $\sigma\in\{0.1,0.5,1\}$ with no systematic growth in $\sigma$, and spatial-only and spatio-temporal variants track each other closely. The transversal layer does not introduce a primitive-specific sensitivity to heterogeneity in this regime; with matched decoder priors, the per-component perturbation is absorbed in the same way as in memory.

\subsection{Lattice Surgery}
Lattice surgery, introduced in \S\ref{subsec:lattice surgery}, realizes joint logical parity measurements ($M_{XX}$ or $M_{ZZ}$) by transiently expanding the stabilizer set across two patches along an ancilla bridge of length $L$ for $t_{\mathrm{merge}}$ rounds, then contracting back for $t_{\mathrm{post}}$ post-surgery rounds; its multi-patch structure exposes both storage and merge-boundary fault locations to the noise model. We study how its performance changes with bridge length, Pauli bias, measurement bias, and non-uniform noise.

\subsubsection{Bridge length under uniform depolarizing noise}

\begin{figure}
    \centering
    \includegraphics[width=1\linewidth]{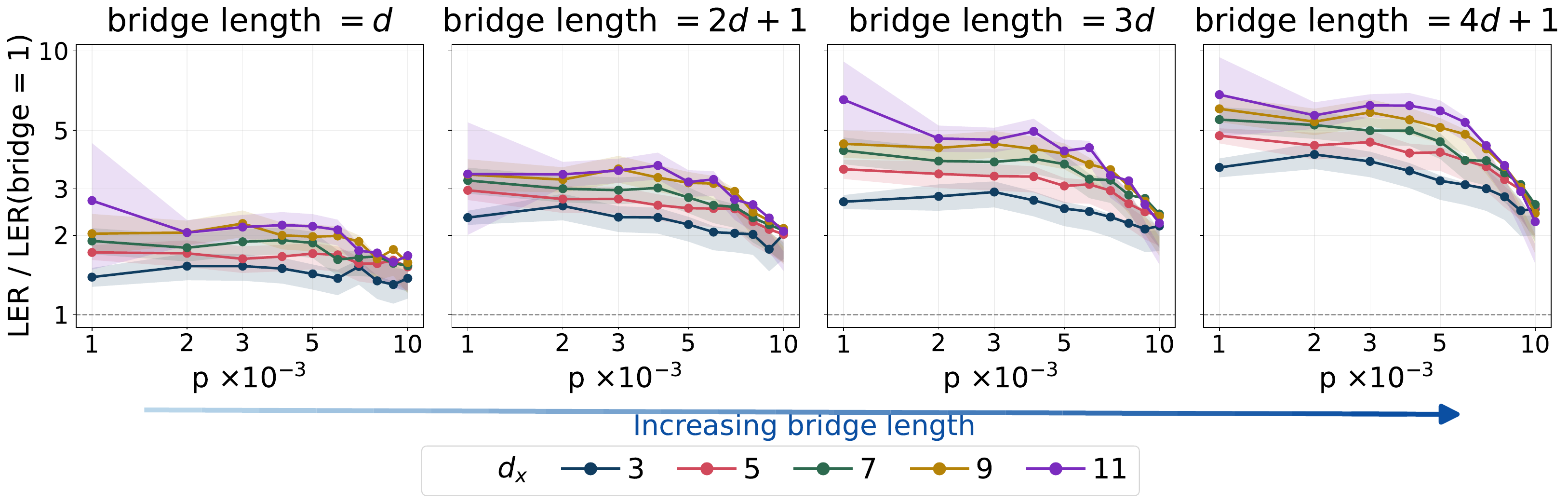}
    \caption{Effect of ancilla bridge length on lattice surgery under uniform depolarizing noise. Each panel shows the relative logical error rate corresponding shortest-bridge experiment with $L=1$, isolating the overhead introduced by increasing the ancilla bridge length.}
    \label{fig:LS_vary_anc_len}
\end{figure}

Figure~\ref{fig:LS_vary_anc_len} first evaluates the effect of the ancilla bridge length under uniform depolarizing noise. Increasing the bridge length introduces additional physical qubits, gates, idling locations, and measurement outcomes in the merge region. As expected, the logical error rate increases with bridge length relative to the shortest-bridge baseline.

The increase is moderate over the bridge lengths studied here, but it is systematic. This confirms that routing overhead increases number and location of physical fault opportunities that can contribute to the joint-parity failure. To isolate the effect of structured noise in the remaining experiments, we fix the bridge length to the shortest layout.

\subsubsection{Z-biased noise}

\begin{figure}
    \centering
    \includegraphics[width=0.99\linewidth]{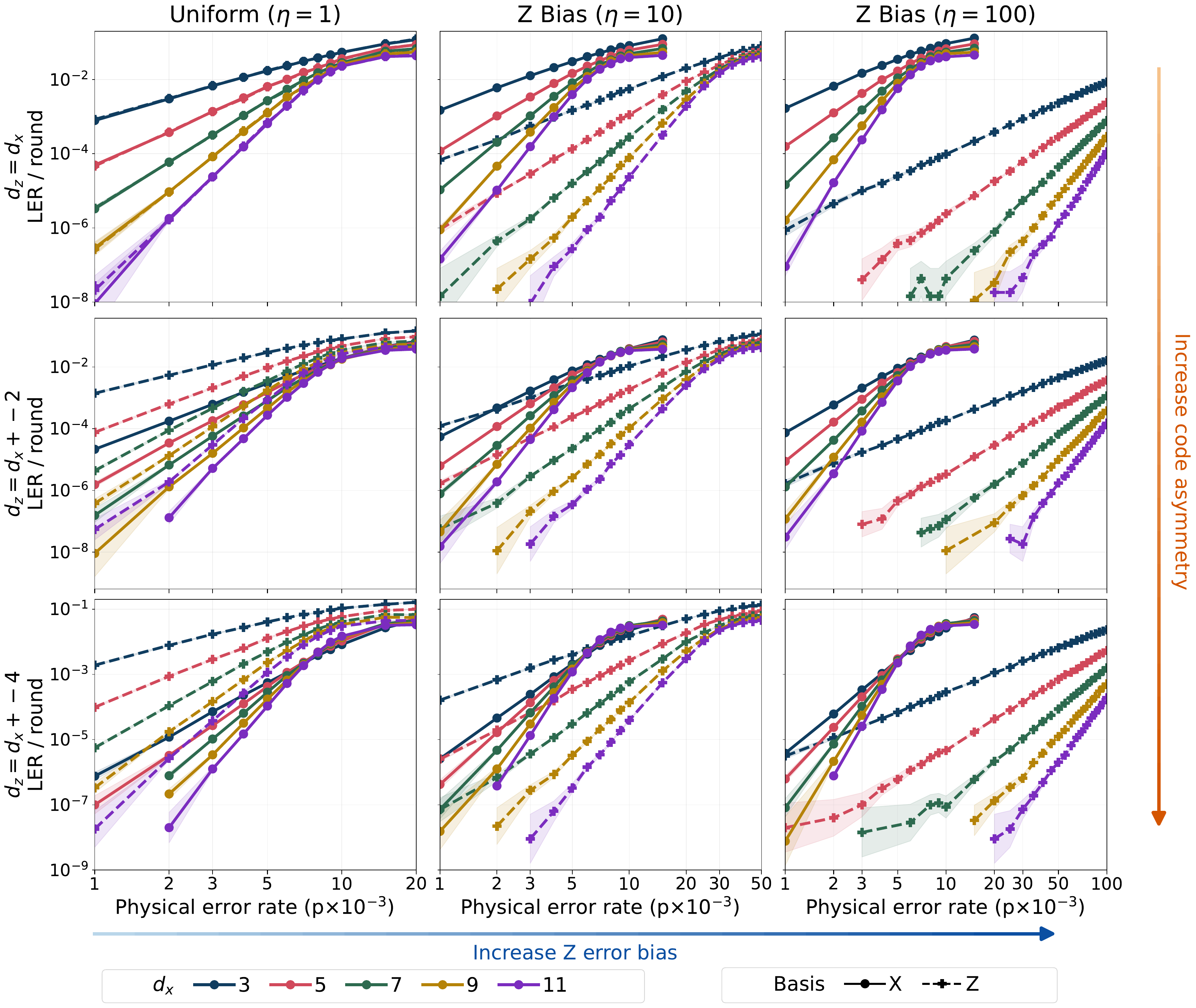}
    \caption{Logical lattice surgery ($M_{ZZ}$) under uniform depolarizing and $Z$-biased noise for the single-ancilla construction. LER per round versus physical error rate; columns sweep $\eta\in\{1,10,100\}$ and rows increase patch asymmetry by enlarging $d_Z$. Under $Z$ bias, rectangular patches with $d_Z>d_X$ reduce the logical error rate by up to an order of magnitude at $\eta=100$, and the geometry advantage appears at lower physical error rates than in memory because the merge--split boundary contributes additional bias-aligned fault locations whose effect is also governed by $d_Z$. Lattice surgery is therefore even more rewarding of anisotropic geometry than memory is---provided the merge boundary is oriented with the suppressed-error axis.}
    \label{fig:ls-z-biased}
\end{figure}



Figure~\ref{fig:ls-z-biased} reports the logical error rate of lattice surgery under uniform and $Z$-biased noise for the single-ancilla layout. As in memory, increased $Z$ bias was mitigated by patches with $d_Z>d_X$, because the dominant $Z$ channel is aligned with the code distance receiving extra protection. Unlike memory, however, lattice surgery is a multi-patch primitive whose logical fidelity is set not only by storage-like suppression but also by the quality of the joint-parity measurement along the merge region. The merge region introduces additional fault locations that contribute directly to the joint-parity measurement and are also governed by $d_Z$. As a result, the threshold behavior is more strongly affected in lattice surgery than in memory. This is visible in the middle and right columns of Fig.~\ref{fig:ls-z-biased}, where the separation between threshold of two logical channels becomes more pronounced as $\eta$ increases.

\subsubsection{Measurement-biased noise}

\begin{figure}[t]
    \centering
    \includegraphics[width=0.99\linewidth]{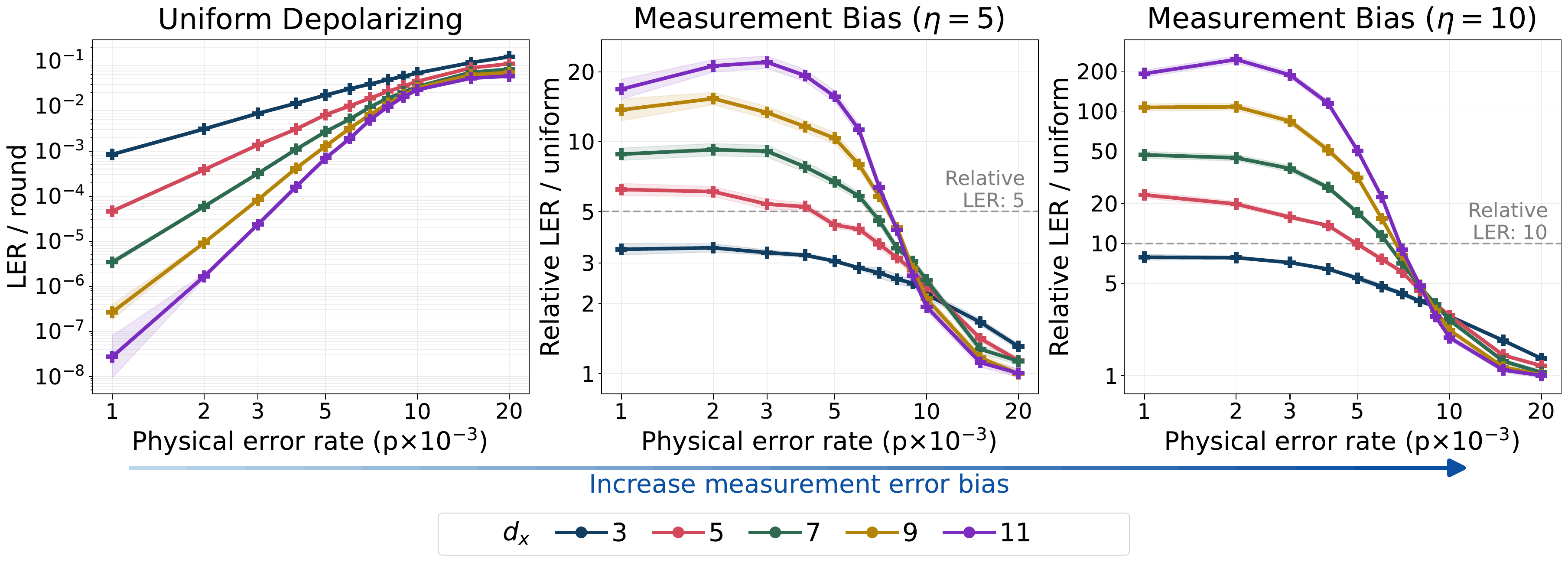}
    \caption{Absolute logical-lattice-surgery LER per round versus round count on square patches with merge distance equal to the round count. Columns compare uniform depolarizing noise with measurement-biased noise at $\eta\in\{5,10\}$. The round count minimizing the LER shifts upward as the measurement bias increases, indicating that lattice surgery benefits from extra temporal redundancy under readout-limited hardware---at a rate that grows with the bias.}
    \label{fig:ls-meas-abs-rounds}
\end{figure}

\begin{figure}[t]
    \centering

    \begin{subfigure}[t]{\linewidth}
        \centering
        \includegraphics[width=0.99\linewidth]{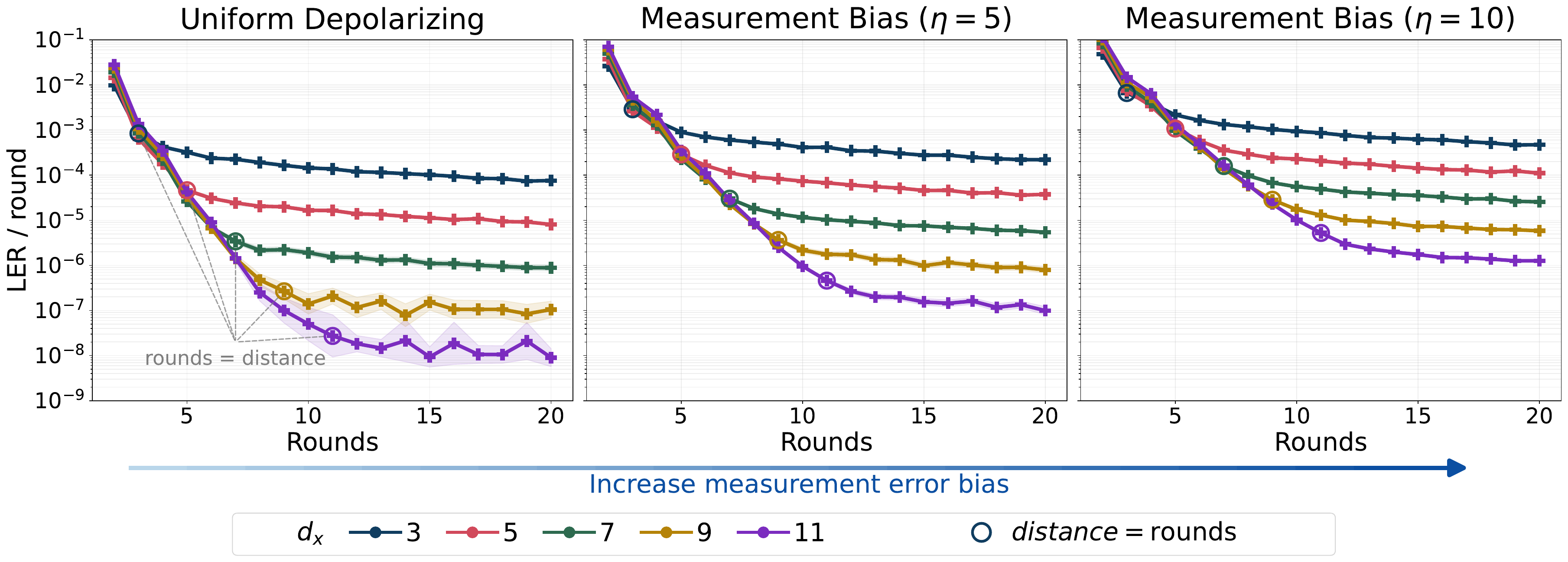}
        \caption{Varying physical error rate at rounds $=d$.}
        \label{fig:ls-meas-relative-p}
    \end{subfigure}

    \vspace{0.5em}

    \begin{subfigure}[t]{\linewidth}
        \centering
        \includegraphics[width=0.7\linewidth]{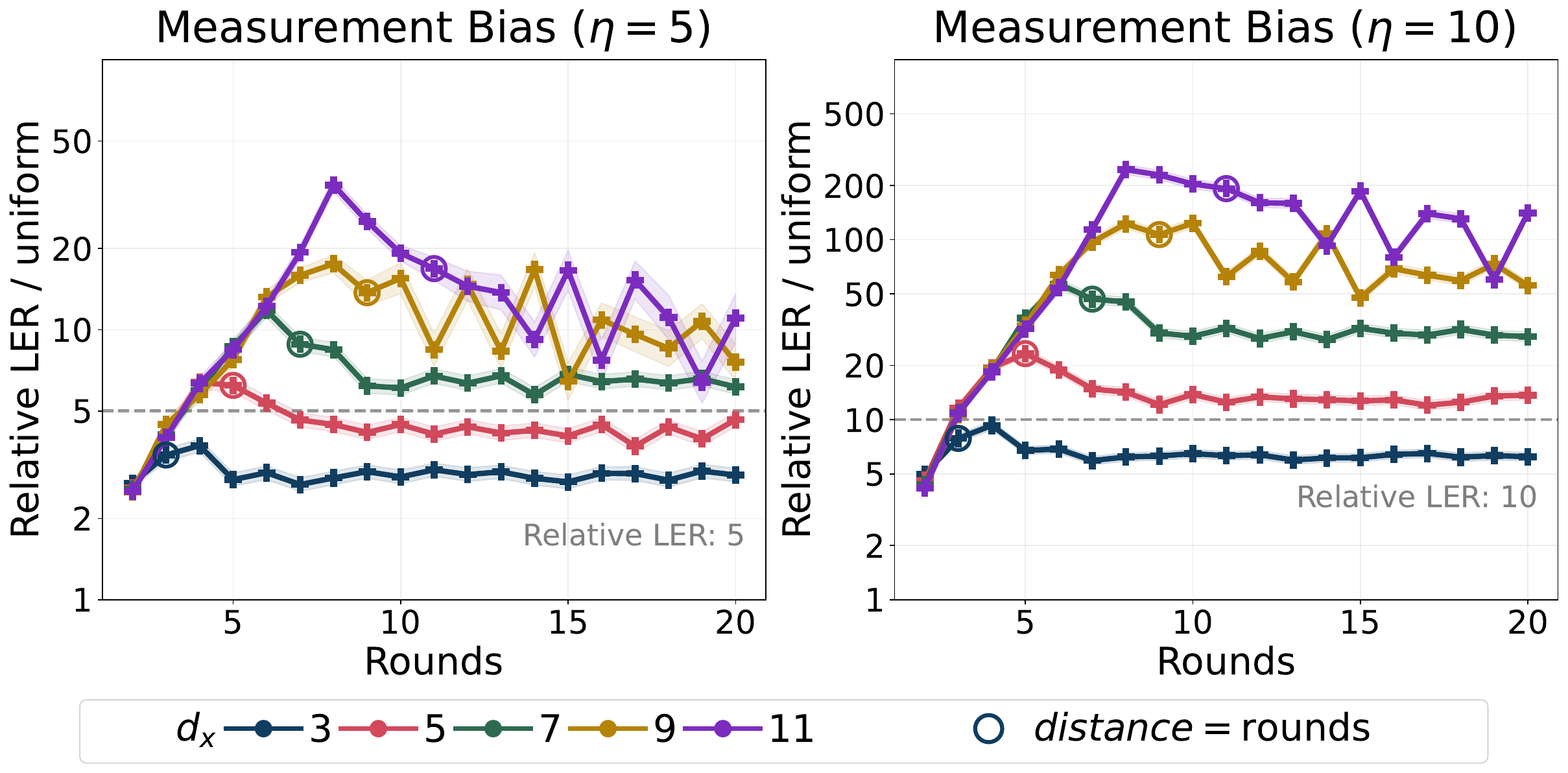}
        \caption{Varying rounds at fixed $p$. Stronger measurement bias keeps the curve above the baseline longer and delays saturation.}
        \label{fig:ls-meas-relative-rounds}
    \end{subfigure}

    \caption{Logical lattice surgery under measurement-biased noise, normalized to the uniform-depolarizing baseline on square patches. Panels isolate the two axes along which the bias penalty evolves---physical error rate and round count---at $\eta\in\{5,10\}$. Relative overhead grows systematically with the bias, peaks at intermediate physical error rates, and decays slowly with added rounds. Measurement-dominated platforms therefore require round schedules and patch geometries to be co-optimized: neither alone suffices to recover near-uniform logical performance.}
    \label{fig:ls-meas-relative-combined}
\end{figure}

The lattice-surgery measurement-bias results make the temporal-overhead story particularly explicit. Figure~\ref{fig:ls-meas-abs-rounds} shows the absolute LER per round versus the number of syndrome-extraction rounds, and the round count minimizing the LER shifts upward as the measurement bias increases---from $t^\star\approx d$ at $\eta=1$ to visibly larger values at $\eta=5$ and $\eta=10$. Figures~\ref{fig:ls-meas-relative-p} and~\ref{fig:ls-meas-relative-rounds} normalize the same effect to the uniform-depolarizing baseline. Together they show that measurement-biased hardware does not simply rescale the logical error rate; it changes the round schedule at which the primitive is most effective. This is consistent with the intuition that lattice surgery needs enough repeated syndrome information to overcome unreliable readout, while each extra round still carries its own exposure to circuit faults---yielding a non-trivial optimum that drifts with the bias.

\subsubsection{Non-uniform noise}
\begin{figure}
    \centering
    \includegraphics[width=0.9\linewidth]{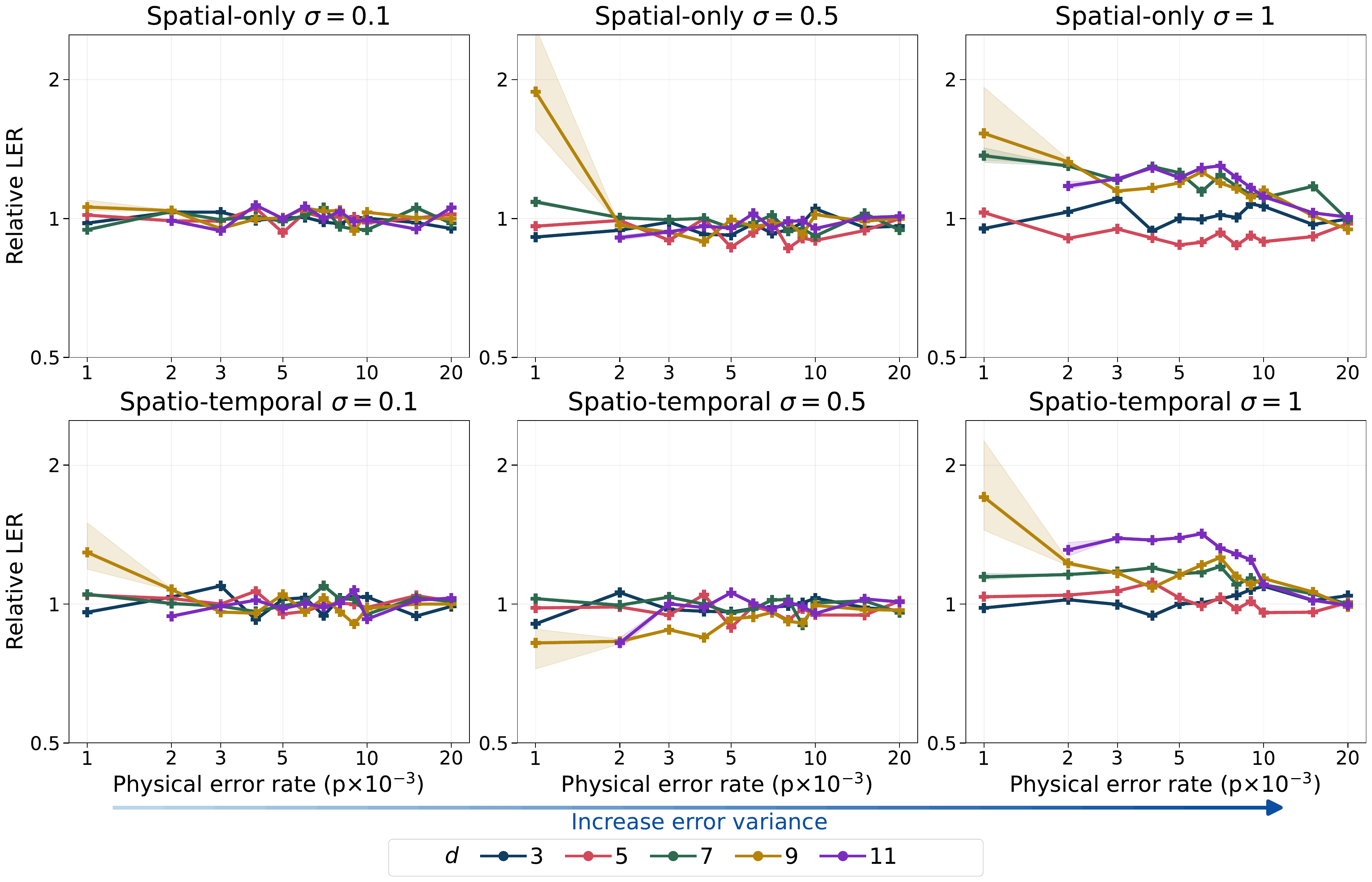}
    \caption{Logical lattice surgery under non-uniform noise, normalized to the uniform-depolarizing baseline on square patches in the $Z$-basis. Heterogeneity strength $\sigma$ scales a Gaussian perturbation applied to the per-component error rate.}
    \label{fig:ls-nonuniform}
\end{figure}

Figure~\ref{fig:ls-nonuniform} shows how lattice surgery responds to heterogeneity across qubits alone (top row) and across qubits and rounds together (bottom row). Relative LER stays within roughly $\pm 20\%$ of the uniform-depolarizing baseline across $\sigma\in\{0.1,0.5,1\}$, with the largest deviations at low $p$ decaying toward unity as $p$ approaches threshold; spatial-only and spatio-temporal variants track each other closely.

\subsection{Logical Phase Gate}
The phase ($S$) gate, introduced in \S\ref{subsec:S}, is implemented via lattice surgery against a $|Y\rangle$-basis measurement on an ancilla qubit, inheriting the merge-region sensitivities of plain lattice surgery while adding a $|Y\rangle$ transition round from~\cite{Gidney2024inplaceaccessto}. We fix $t_{\mathrm{pre}}=0$ throughout, so the only round-count knobs are the merge length $t_{\mathrm{merge}}$ and the post-transition padding $t_{\mathrm{boundary}}$. The gate is not separately analyzed under $Z$-biased noise: logical $S$ commutes with $\bar{Z}$, so a $Z$-dominant fault distribution passes through the gate unchanged rather than being redistributed across bases as in the transversal Hadamard, and the underlying $M_{ZZ}$ surgery already characterizes the only $Z$-sensitive component of the protocol. Measurement bias instead stresses the boundary readout on which the merge depends, while spatial and spatio-temporal heterogeneity probe whether a single mis-calibrated qubit along the merge boundary can dominate the joint-parity outcome. We use the same matched $(d,\,\mathrm{bridge\_length}=1,\,p,\,\text{total round count})$ baseline as the rest of the section and report relative LER against the uniform-depolarizing reference.

\subsubsection{Measurement-biased noise}
\begin{figure}
    \centering
    \includegraphics[width=1\linewidth]{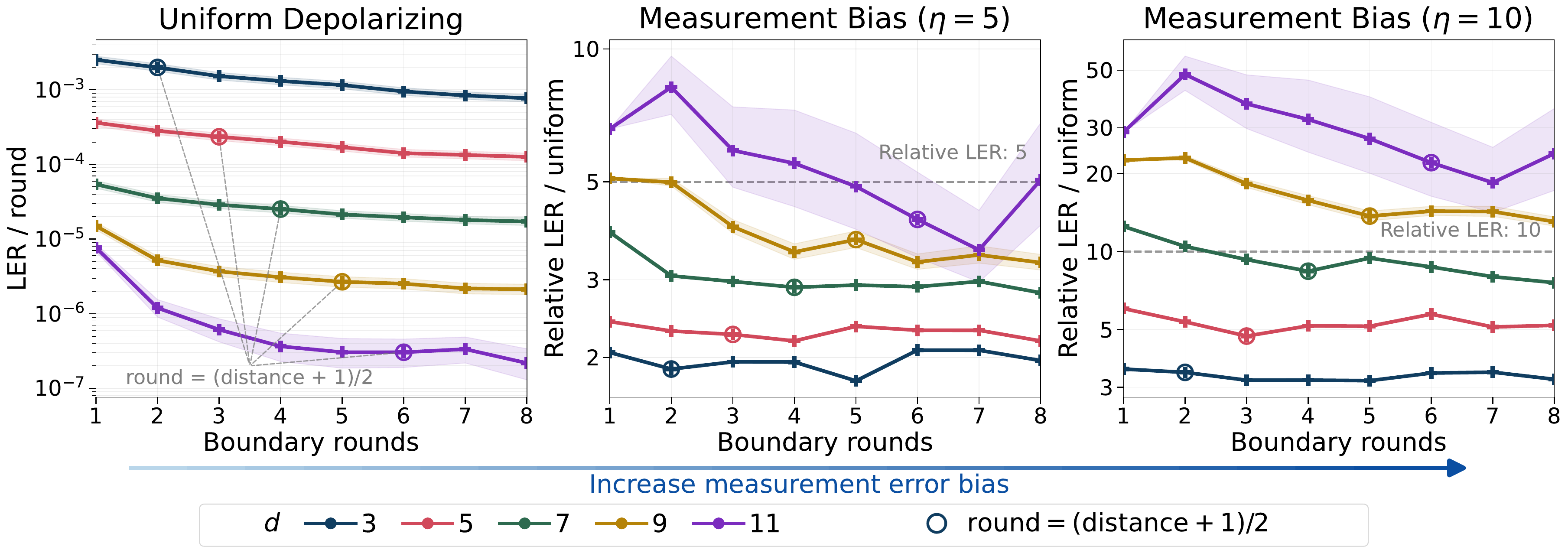}
    \caption{Logical phase gate under measurement-biased noise: relative LER versus boundary-round count $t_{\mathrm{boundary}}$ at $t_{\mathrm{pre}}=0$, $t_{\mathrm{merge}}=d$, and $p=10^{-3}$, for $\eta \in \{5, 10\}$. The relative LER decreases with $t_{\mathrm{boundary}}$ and saturates near the $(d+1)/2$ prescription of Ref.~\cite{Gidney2024inplaceaccessto}; the effect is stronger at larger $d$ but modest in absolute terms, indicating the transition-round error is not the dominant fault contribution in the protocol.}
    \label{fig:s-meas-vary-round}
\end{figure}

The phase-gate measurement-bias results isolate the cost of the boundary-round schedule used to suppress the $\bar{Y}$ transition-round error. Ref.~\cite{Gidney2024inplaceaccessto} prescribes $t_{\mathrm{boundary}} = (d+1)/2$ as the minimum padding required to absorb that error. Figure~\ref{fig:s-meas-vary-round} sweeps $t_{\mathrm{boundary}}$ at fixed $t_{\mathrm{pre}}=0$, $t_{\mathrm{merge}}=d$, and $p=10^{-3}$, validating this scaling: the relative LER decreases with $t_{\mathrm{boundary}}$ and saturates near the prescribed value. The effect is more pronounced at larger $d$, but its overall magnitude is modest, indicating that the transition-round error is not the dominant fault contribution in the gate-teleportation protocol.

\begin{figure}
    \centering
    \includegraphics[width=1\linewidth]{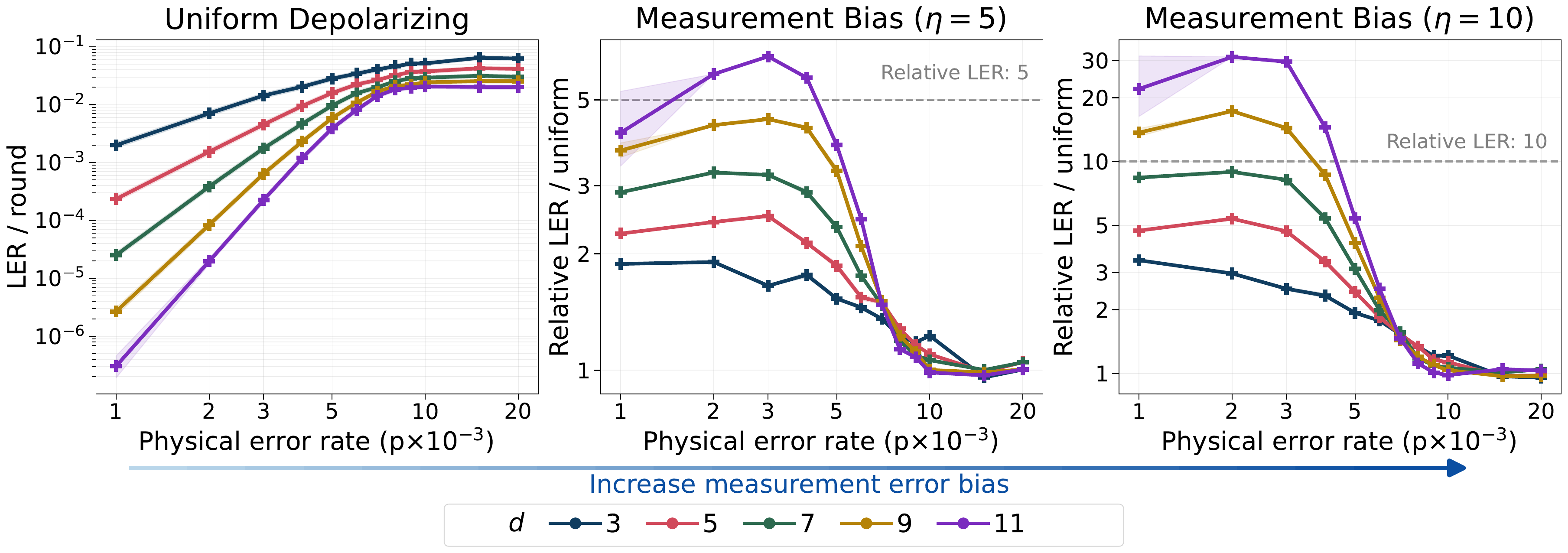}
    \caption{Logical phase gate under measurement-biased noise: relative LER versus physical error rate at $t_{\mathrm{boundary}}=(d+1)/2$, $t_{\mathrm{pre}}=0$, and $t_{\mathrm{merge}}=d$, for $\eta \in \{5, 10\}$. As in memory, relative overhead grows with the bias, is amplified at larger code distances, and is suppressed near threshold where the absolute LER saturates.}
    \label{fig:s-meas-vary-p}
\end{figure}

Figure~\ref{fig:s-meas-vary-p} fixes $t_{\mathrm{boundary}}=(d+1)/2$, $t_{\mathrm{pre}}=0$, and $t_{\mathrm{merge}}=d$ and sweeps the physical error rate at $\eta \in \{5, 10\}$. The phase gate exhibits the same qualitative pattern as memory under measurement bias: relative LER grows with $\eta$, is largest at intermediate $p$ and at higher code distances, and is suppressed near threshold where the absolute LER saturates and structured-noise overhead is masked by overall code degradation.

\subsubsection{Non-uniform noise}
Applying the non-uniform noise model of \S\ref{subsec:nonuniform_noise}, the phase gate exhibits the same qualitative response as memory, transversal Hadamard, and lattice surgery: relative LER stays close to the uniform-depolarizing baseline at $\sigma \in \{0.1, 0.5, 1\}$, with little separation between spatial-only and spatio-temporal disorder. This reinforces the broader observation that the rotated surface code is comparatively insensitive to per-qubit calibration heterogeneity across the primitives studied here.

\begin{figure}
    \centering
    \includegraphics[width=0.9\linewidth]{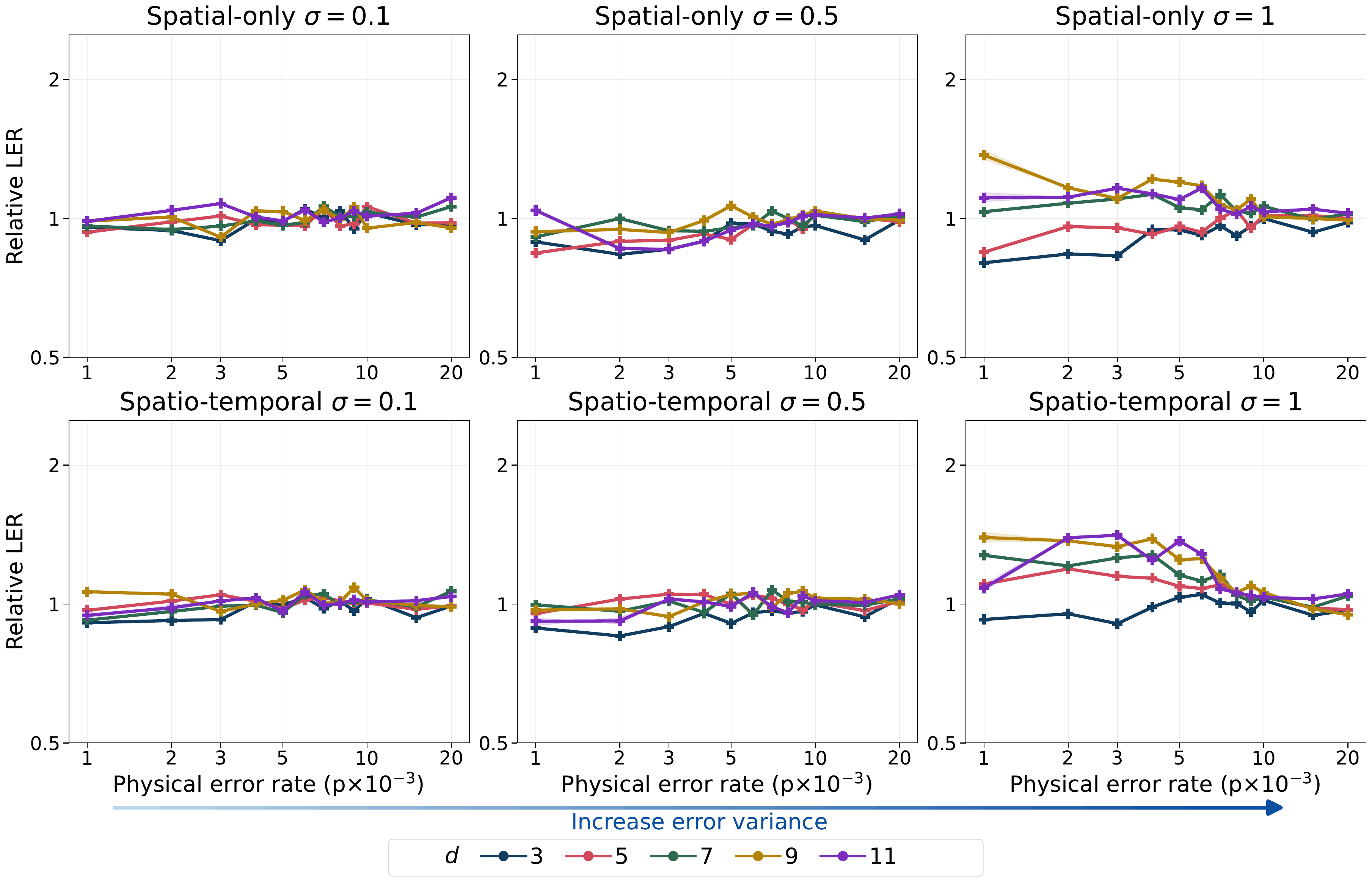}
    \caption{Logical phase gate under non-uniform noise, normalized to the uniform-depolarizing baseline. Rows compare spatial-only and spatio-temporal disorder; columns sweep heterogeneity strength $\sigma\in \{0.1, 0.5, 1\}$. Relative LER remains within a small factor of unity across the surveyed regime, and the spatial-only and spatio-temporal variants track closely.}
    \label{fig:s-nonuniform}
\end{figure}

%% file: sections/discussion.tex
\section{Discussion and Future Directions}
\label{sec:discussion}

The evaluation section covers a structured but finite slice of the hardware-motivated noise design space. The value of \sol{} is not in having exhaustively mapped that space, but in making the process of extending such a study tractable. We conclude by describing how the framework can be pushed beyond the cases shown here.

\paragraph{Extensibility to other decoders and simulators.} All circuits produced by \sol{} are emitted in \textsc{Stim}~\cite{gidney2021stim} format, and logical observables and detector annotations are bundled with each circuit so that the corresponding Detector Error Model (DEM) can be generated without additional tooling. Any decoder or simulation environment that consumes this format---for example \textsc{PyMatching}~\cite{Higgott2025sparseblossom}, Tesseract~\cite{beni2025tesseractdecoder}, tsim~\cite{tsim2026} and \textsc{StabSim}~\cite{garner2025stabsim}---can be swapped in without modifying the upstream generator. The evaluations reported here therefore serve as one concrete instantiation; the same circuits drive correlated-matching, maximum-likelihood, belief-propagation, and learning-based decoders equally, enabling direct cross-decoder comparisons under identical noise assumptions.

\paragraph{Extensibility to other code families.} Although we focus on the rotated surface code, the noise-modeling interface and the \textsc{Stim}-format output are code-agnostic. Bias-tailored variants such as the XZZX surface code~\cite{tuckett2018ultrahigh, tuckett2019tailoring}, color codes~\cite{landahl2011fault}, and qLDPC constructions can be plugged in by supplying their stabilizer generators and logical-operator definitions; the noise channels, spatial/temporal assignment, and idling accounting described in \sol{} then apply identically. This makes \sol{} a natural vehicle for head-to-head comparisons between code families under matched, hardware-motivated noise---an important direction given the growing number of candidate codes.

\paragraph{Toward co-design.} Taken together, the portability of the output format, the extensibility of the primitive and noise generators, and the reproducibility of the standardized pipeline position \sol{} as an infrastructure layer for hardware--software co-design: the same configuration files that capture a device's calibration profile can drive evaluations across primitives, codes, and decoders, so that design decisions---patch geometry, syndrome-extraction schedules, routing of merge boundaries, decoder choice---can be revisited as hardware characterization data evolves.